\documentclass[final, 11pt]{article}
\pdfoutput=1
\usepackage{MRnotes}

\newcommand{\RNAdS}[1]{Reissner-Nordstr\"om-AdS$_{#1}$}

\catcode`,\active

\catcode`\,12

\newcommand*\pFqRegskip{8mu}
\catcode`,\active
\newcommand*\pFqReg{\begingroup
        \catcode`\,\active
        \def ,{\mskip\pFqRegskip\relax}%
        \dopFqReg
}
\catcode`\,12
\def\dopFqReg#1#2#3#4#5{%
        {}_{#1}\textbf{F}_{#2}\biggl(\genfrac..{0pt}{}{#3}{#4};#5\biggr)%
        \endgroup 
        }

\catcode`,\active

\catcode`\,12

\newcommand{\ann}{\mathscr{M}}   % Markovianity index
\newcommand{\sen}[1]{\varphi_{_{#1}}} % designer scalar
\newcommand{\wk}{\bar{k}}
\newcommand{\bwt}{\mathfrak{w}}   % dimless freq
\newcommand{\bqt}{\mathfrak{q}}    % dimless momentum
\newcommand{\bk}{\vb{k}}  % spatial momentum 
\newcommand{\bx}{\vb{x}}   % space coordinates
\newcommand{\bpt}{\mathfrak{p}} % lightcone momenta
\newcommand{\ctor}{\zeta}  % the mock tortoise coordinate introduced in 2004
\newcommand{\ri}{\varrho}   % inverse radial (primarily in appendices) 
\newcommand{\Dz}{\mathbb{D}}  % The TR derivative
\newcommand{\ai}{\alpha}   % Vector pol label for harmonics
\newcommand{\PHE}{\Phi_{_\text{E}}}
\newcommand{\PHO}{\Phi_{_\text{O}}}
\newcommand{\PHW}{\Phi_{_\text{W}}}
\newcommand{\AGR}{\mathscr{A}}

\newcommand{\Lk}{\Lambda_k}

\newcommand{\MV}{\mathscr{V}}
\newcommand{\MZ}{\mathscr{Z}}  % scalar sector
\newcommand{\MX}{\mathscr{X}}  % vector sector
\newcommand{\MY}{\mathscr{Y}}  % tensor sector

\newcommand{\In}{\text{\tiny{in}}}     % ingoing label 
\newcommand{\Out}{\text{\tiny{out}}}  % Outgoing solution
\newcommand{\nB}{n_{_B}}   % BE factor
\newcommand{\Gin}[1]{G^{{#1}}_\In}     % Ingoing propagator
\newcommand{\Kin}[1]{K_{_{#1}}}      % the ingoing bdy correlation propagator 
\newcommand{\Gout}[1]{G^{{#1}}_\Out}   % Outgoing propagator
\newcommand{\gfn}[1]{\mathfrak{G}_{_{#1}}}

\newcommand{\sdc}{\mathfrak{S}_{_Q}}
\newcommand{\RQ}{r_{_Q}}
\newcommand{\bRQ}{\mathfrak{C}}
\newcommand{\BQTv}{\mathfrak{p}_v}
\newcommand{\BQTs}{\mathfrak{p}_s}
\newcommand{\Zd}{\mathsf{Z}}
\newcommand{\Vd}{\mathsf{V}}
\newcommand{\VVd}{\mathfrak{V}_\Vd}
\newcommand{\VZd}{\mathfrak{V}_\Zd}

\newcommand{\skR}{\text{\tiny R}}
\newcommand{\skL}{\text{\tiny L}}

\newcommand{\JF}{J_{_\text{F}}}
\newcommand{\JP}{J_{_\text{P}}}
\title{Holographic thermal correlators: A tale of Fuchsian ODEs and integration contours}
\author[a]{R. Loganayagam,}  
\author[b]{ Mukund Rangamani,}
\author[b]{ Julio Virrueta}

\affiliation[a]{
    International Centre for Theoretical Sciences (ICTS-TIFR), \\ 
    Tata Institute of Fundamental Research, Shivakote, Hesaraghatta, Bangalore 560089, India.}
 
\affiliation[b]{
    Center for Quantum Mathematics and Physics (QMAP)\\
    Department of Physics \& Astronomy, University of California, Davis, CA 95616 USA}
\emailAdd{nayagam@icts.res.in}
\emailAdd{mukund@physics.ucdavis.edu}
\emailAdd{jvirrueta@ucdavis.edu}

\abstract{ 
We analyze real-time thermal correlation functions of conserved currents in holographic field theories using the grSK geometry, which provides a contour prescription for their evaluation. We demonstrate its efficacy, arguing that there are situations involving components of conserved currents, or derivative interactions, where such a prescription is, in fact, essential. To this end, we first undertake a careful analysis of the linearized wave equations in AdS black hole backgrounds and identify the branch points of the solutions as a function of (complexified) frequency and momentum. All the equations we study are Fuchsian with only regular singular points that for the most part are associated with the geometric features of the background. Special features, e.g., the appearance of apparent singular points at the horizon, whence outgoing solutions end up being analytic, arise at higher codimension loci in parameter space. Using the grSK geometry, we demonstrate that these apparent singularities do not correspond to any interesting physical features in higher-point functions. We also argue that the Schwinger-Keldysh collapse and KMS conditions, implemented by the grSK geometry, continue to hold even in the presence of such singularities. For charged black holes above a critical charge, we furthermore demonstrate that the energy density operator does not possess an exponentially growing mode, associated with `pole-skipping', from one such apparent singularity. Our analysis suggests that the connection between the scrambling physics of black holes and energy transport has, at best, a  limited domain of validity. 
}

\begin{document}
\maketitle

%%%%%%%%%%%%%%%%%%%%%%%%%%%%%%%%%%%%%%%%%%%%%%%%%%%
 
%~~~~~~~~~~~~~~~~~~~~~~~~~~~~~~~~~~~~~~~~~~~~~~~
\section{Introduction}
\label{sec:intro}
%~~~~~~~~~~~~~~~~~~~~~~~~~~~~~~~~~~~~~~~~~~~~~~

The study of linearized  perturbations of black holes, initiated in ~\cite{Vishveshwara:1970cc,Press:1971wr}, is relevant for a broad class of physical questions. These range from astrophysical signals of black hole horizon ring-down, to mathematical relativity questions of black hole stability, and holography~\cite{Horowitz:1999jd}. The linear wave equations have therefore been analyzed by many authors over the years.\footnote{ In addition to the classic reviews of the subject~\cite{Kokkotas:1999bd,Berti:2009kk}, a recent review~\cite{Hatsuda:2021gtn} might prove helpful in the context of our discussion.} Our interest is in the context of holography, especially in the analytic  structure of the solutions, and how this informs boundary correlation functions, in particular higher-point functions.

The context for our analysis is the computation of real-time thermal correlators in a strongly coupled holographic field theory. We will focus on Schwinger-Keldysh correlators, which have been argued to be computed on a particular complexified spacetime in~\cite{Glorioso:2018mmw} (we refer to the resulting geometry as the grSK spacetime). As it stands, their prescription  can be viewed as a particular contour choice for the computation of bulk Witten diagrams, as elaborated in ~\cite{Chakrabarty:2019aeu,Jana:2020vyx,Loganayagam:2022zmq}.
The virtue of having a contour prescription is that one can fold in the information about the nature of local behaviour into the contour integral. This turns out to be especially important when considering correlation functions of conserved currents where, as we shall see, the bulk vertices involve functions with singularities which potentially interfere with the contour of integration. In turn, these new singularities could, in principle, break the general arguments advanced in the above references that correlators computed in the grSK spacetime satisfy the general properties required of Schwinger-Keldysh thermal correlators. 

To be clear, the analysis of real-time correlation functions in the holographic context has a long and rich history. The original prescription for computing two-point functions was given in~\cite{Son:2002sd} and justified in~\cite{Herzog:2002pc}. The logic here was to argue that the retarded correlators, by virtue of boundary causality, should involve infalling boundary conditions on the future horizon and suitable boundary sources controlled by non-normalizable modes at the AdS asymptopia.\footnote{ Equivalently, the infalling modes are naturally the analytic solutions for the perturbation equations in a suitable basis.} 
The boundary two-point function was obtained as the ratio of the normalizable mode to the non-normalizable mode for such solutions. Much of the subsequent literature in applications of AdS/CFT to compute thermal real-time observables, spectral functions, fluid-gravity correspondence etc., has relied on this prescription. 

While the prescription of~\cite{Son:2002sd} captures correctly the retarded observables, owing to the presence of a single boundary condition on the asymptopia, it does not naturally lend itself to computing fluctuations (one could, however, get these by demanding that the fluctuation-dissipation theorems hold at the Gaussian order). A more natural prescription therefore was the proposal of~\cite{Skenderis:2008dh,Skenderis:2008dg} who argued that one should take the boundary Schwinger-Keldysh contour seriously. In particular, they posited that the bulk spacetime dual to such contours should be constructed by piecewise gluing Lorentzian and Euclidean geometries across codimension-1 bulk surfaces. Within this framework,~\cite{vanRees:2009rw} demonstrated the naturalness of the infalling prescription of~\cite{Son:2002sd}. Taking this into account, one can indeed compute (higher-point) correlators with general Schwinger-Keldysh time-ordering, as was done already a decade ago in~\cite{Arnold:2011ja,Arnold:2011hp}. More recent efforts in this direction include~\cite{Botta-Cantcheff:2015sav,Botta-Cantcheff:2017qir} for non-thermal excited states, and~\cite{Pantelidou:2022ftm} who computed thermal 3-point functions.

One might thus ask, what does the prescription of~\cite{Glorioso:2018mmw} buy us, apart from perhaps the elegance associated with writing the result in terms of a contour integral? For instance, using this prescription it was demonstrated that a general correlation function of a class of boundary operators could be written as an integral over a single copy of the black hole exterior, of an integrand that is a (multiple) discontinuity~\cite{Jana:2020vyx,Loganayagam:2022zmq} (see also~\cite{Loganayagam:2022xyz}). However, certain assumptions\footnote{
    In~\cite{Loganayagam:2022zmq} it was assumed that the bulk vertex functions do not introduce any additional singularities; all singularities in the integrand can be traced back to the linearized solutions of the wave equation.} 
were made in the process of deriving these statements. It transpires that for a class of operators, the bulk vertices for the corresponding holographic fields involves non-analytic vertices. Having a contour prescription allows one to disambiguate these situations, and allows one to  argue in its favour. 

A simple example which explains the issue is a cubic interaction of two bulk-fields (say $\phi$ and $\chi$) involving derivatives, e.g., $\chi\, \nabla_A \phi\, \nabla^A \phi$. In this case, the interaction vertex written in a basis adapted to the analytic infalling modes has a pole at the location of  the horizon. If we were to say that the prescription involves gluing together two copies of the black hole exterior along the codimension-1 null hypersurface (the horizon) we have a jump discontinuity in the (null) extrinsic curvature, which requires regulating. The contour integral provides a natural regulator, giving a clean prescription for the boundary correlator. 

Another situation where one encounters something interesting corresponds to the computation of correlation functions of the boundary energy density operator. Now the integrands involve a function $\Lk$ of geometric data and spatial momenta $k$,~\eqref{eq:Lkdef}, whose zeros $r_k$ potentially pinch the integration contour for certain values of $k \in \mathbb{C}$.\footnote{ 
    The function has $d-2$ complex zeros on a circle of radius $r_k$. One of these  lies between the horizon and the boundary, at a real value of the radial coordinate, for $\abs{k} \in [k_\star,\infty)$ and $\arg{k^2}=\pi$. The lower bound $k_\star$ is the special momentum when the root is at the horizon. \label{fn:rayLkS}} 
This behaviour turns out to be related to explorations in the literature regarding the contribution of the energy density operator to the out-of-time-order correlator capturing chaotic and scrambling dynamics~\cite{Blake:2017ris} (see also~\cite{Haehl:2018izb} for an analysis in 2d CFTs). From a gravitational perspective the phenomenon of interest dubbed `pole-skipping' in the aforementioned work, refers to loci in the complex frequency $\omega$ and momentum $k$ space where the naive expectation of the correlator having a pole turns out to be unfounded. As observed originally in~\cite{Grozdanov:2017ajz} and elaborated upon in~\cite{Blake:2018leo}, in the holographic context for neutral black holes, the phenomenon occurs at a codimension-2 locus in the $(\omega,k)$ space, at  $(\omega_\star,k_\star) =2\pi i\, T (1, \sqrt{\frac{2\,(d-1)}{d}})$. The real space profile of the solution is that of an exponentially growing mode at the rate set by the maximal Lyapunov exponent $\lambda_L = 2\pi i \, T$, damping out spatially with a rate set by the butterfly velocity $v_B =\sqrt{\frac{d}{2\,(d-1)}} $~\cite{Shenker:2013pqa}. 

The phrase \emph{apparent quasinormal modes} captures the  physical aspects of this behaviour better~\cite{Loganayagam:2022zmq}, for, as first noticed in~\cite{Blake:2018leo} and analyzed in various other contexts in~\cite{Grozdanov:2019uhi,Blake:2019otz,Natsuume:2019xcy}, at these loci the two-point correlator is ambiguous. From a mathematical perspective, at these particular values of the parameters the differential equations governing linearized fluctuations no longer have a regular singular point at the horizon. Rather the horizon becomes an apparent singularity. Physically, this means that the outgoing mode, which is generically non-analytic,\footnote{Recall that we choose to make the ingoing mode manifestly analytic by working in infalling coordinates.} stops being so. Consequently, there is no branch cut in the grSK geometry and thus the retarded two-point function becomes ill-defined. Per se, this is not a problem, happening as it does in a set of measure zero. However, given that this happens when the function $\Lk$ has a zero at the horizon (which  a-priori could have led to an irregular singular point), one might ask if there are other regions where the poles of this function can interfere with the grSK contour.

We address this question and show that apart from the horizon, which is relevant for correlators of generic primaries, global currents, and the transverse tensor and vector polarizations of the stress tensor~\cite{Ghosh:2020lel}, the only other point where the function changes the character of the solution corresponds to the asymptotic boundary, which owes to the presence of soft modes (as was understood in~\cite{He:2022jnc}). In summary, we establish that for all bosonic wave equations of interest in the \SAdS{d+1} background, the only singular points of the linearized equations are at the asymptopia, black hole singularity, and at the horizon (and other roots of the metric function $f$). 

While it is useful to establish that the solution to the wave equation corresponding to physical (gauge-invariant) gravitational fluctuations is well-behaved at other radial positions, it is not quite sufficient for the analysis of correlation functions.  First note that the energy density operator couples to all the operators of the CFT; in the bulk this translates to the statement that the scalar graviton polarizations have a non-trivial vertex with all primaries (and contributes to $\expval{T_{\mu\nu} \;\mathcal{O}\,\mathcal{O}}$). The corresponding bulk degree of freedom, $\MZ$, a scalar graviton polarization,  satisfies an autonomous differential equation with only an apparent singular point at the roots of the function $\Lk$, and is therefore regular there. The metric functions themselves are determined not only in terms of $\MZ$ and its derivatives, but crucially have a $\Lk$ dependence that diverges as $r-r_k$ (the transformation  between the metric perturbations and $\MZ$ has a simple pole). Should the residue at this pole be non-vanishing for $k$ lying along the ray described in \cref{fn:rayLkS}, we would have a problem satisfying the Schwinger-Keldysh collapse rules and the KMS condition. 

Since this issue only arises upon complexification of momenta, a skeptical reader might ask whether this has any bearing on the physical correlation functions. While we are sympathetic to this view, it seems a bit bizarre that the real-time correlation functions given by the grSK prescription could fail to respect basic consistency conditions along a real codimension-1 surface in the complex $(\omega,k)$ space. Moreover, the universality of the energy density coupling, suggests that the resolution should be sufficiently general. Indeed, we verify by direct computation that not only is $\MZ$ analytic at the zeros $r_k$ of $\Lk$, but so are the metric functions. Therefore, there are no subtleties even in the higher-point correlation functions computed using complexified grSK geometry. In other words, the analysis of these equations, provides a non-trivial consistency check of the grSK contour prescription.

We not only analyze the situation in the \SAdS{d+1} black hole, which has been examined in part in the literature, but also undertake an analogous analysis for the \RNAdS{d+1} black hole. Now one has to contend with additional singular points arising from the charge dependence through the Ohmic radius introduced in~\cite{He:2021jna}. These are benign lying as they do inside the horizon, and thus do not affect the grSK contour. For this charged plasma, however, the longitudinal scalar polarization of stress tensor couples to the longitudinal mode of the charge current. They can be decoupled at the linearized order, leading to an identification of the energy density  and charge diffusion operators ~\cite{He:2022deg} (the corresponding two bulk fields are $\Zd$ and $\Vd$, respectively).

The decoupling again introduces  an analogous function $\Lk$ (still given by~\eqref{eq:Lkdef}). Since the metric function is different, the set of zeroes of $\Lk$ is correspondingly distinct -- it has two sets of zeros on two circles of radii $r_{k1}$ and $r_{k2}$, with phases determined by $k$. Moreover, $\abs{r_{k1}} > r_+$ for small charges,  $\abs{r_{k2}} > r_+$ for large charges, with the swap happening well away from extremality.\footnote{
    The parameterization for \RNAdS{d+1} we use has a charge parameter $Q \in (0,\sqrt{\frac{d}{d-2}}]$ with the upper limit corresponding to extremality. The critical charge where the two sets of roots exchange dominance is lower; it is at $Q_* = \sqrt{\frac{d}{3d-4}}$.
}  The mapping from $\Zd$ and $\Vd$ to the physical metric and gauge field perturbations has poles at the zeros of $\Lk$. The energy density field $\Zd$ has apparent singularities at both sets of roots, but the charge diffusion field $\Vd$, while having an apparent singularity at the $r_{k1}$ set, instead has a simple pole at the roots in the $r_{k2}$ set. For either set of roots we can tune the momentum so that we have a potential pole along the grSK contour. 
However, once again, if we evaluate the metric functions in the vicinity of the $\Lk$ zeros (either set), the metric and gauge field perturbations are manifestly analytic, ensuring that the correlation functions are well-behaved, and respect the Schwinger-Keldysh collapse rules and KMS conditions.

When we further fine-tune the momentum so that one of the roots from the $r_{k1}$ set  lies on the horizon, we find a behaviour analogous to the \SAdS{d+1} case. The horizon can be made an apparent singular point at codimension-2 locus $(\omega_\star,k_\star) =2\pi i\, T (1, \sqrt{\frac{(d-1)\,r_+}{2\pi\, T}})$.\footnote{
    We work in units where $\lads=1$, and thus the radial coordinate has units of energy. $T/r_+$ is dimensionless, and we prefer to emphasize the dependence on the physical temperature measured in horizon size units. 
}
Translating to position domain we again have an exponentially growing mode in time, and damped in space, mimicking the scrambling mode for localized perturbations~\cite{Roberts:2014isa} (see the general expression given in~\cite{Mezei:2018jco}).

Above this critical charge, the aforementioned behaviour shifts to the charge diffusion mode. This however occurs at a different value of frequency, specifically at  $(\omega_\star,k_\star) =2\pi i\, T (-1, \sqrt{\frac{(d-1)\,r_+}{2\pi T}})$. Now one no longer has an exponentially growing mode, even at an apparent singular point of the differential equation. Furthermore, the energy density field does not show any special features; its outgoing mode is non-analytic at the horizon as usual. 

This is perhaps unexpected from the viewpoint of~\cite{Blake:2017ris}, who argued for a general connection between quantum chaotic dynamics encoded in the exponential growth of the out-of-time ordered 4-point function to the energy density 2-point function. This connection fails to hold for charged black holes above a critical charge. While~\cite{Abbasi:2020ykq,Jansen:2020hfd} have earlier analyzed the fluctuations of \RNAdS{4} for connection between conserved current correlators and chaos they appear to have missed this fact. 
We return to what this could imply for maximally chaotic systems in \cref{sec:discuss}. 

The outline of the paper is as follows: In \cref{sec:prelim} we review  the grSK geometry and the contour prescription for evaluating real-time thermal correlators. We also take the opportunity to give a synopsis of the features of second order linear differential equations that we are interested in, reminding the reader of some the standard terminology. This allows us to identify issues that arise in computation of real-time correlation functions using the grSK geometry. In particular, we highlight several potential failure points of the prescription, which we systematically explore in the sequel.

In~\cref{sec:bhGreen} we summarize the upshot of our analysis of wave equations for the Green's functions necessary for the grSK contour. Throughout this work, we assume that our background geometry has time and spatial translational symmetries along the boundary (i.e., CFT) directions, which allows us to work in Fourier domain; our parameters also comprise then the frequencies and momenta. We discuss both the \SAdS{d+1} black hole and the \RNAdS{d+1} black hole. Special focus will be on the fields dual to conserved currents, which have an intricate behaviour.  Having established the analytic structure of the solutions of the perturbations, we then turn  in \cref{sec:intvertex} to the computation of Witten diagrams using the grSK geometry. Here we discuss several scenarios: interaction terms where vertex functions have poles at the horizon, the apparent singularities in the energy density operator, and finally comment on the observations regarding `pole-skipping' behaviour in Schwinger-Keldysh correlation functions. Some of these features are explicitly exemplified by a computation of a three-point function in a simple setting (derivative interaction of two scalar fields in the BTZ geometry).

To keep the main text simple, we have relegated various facts, and some detailed derivations to appendices.  For completeness, we collate all the differential equations we analyze in \cref{sec:eqcompilation} and some formulae relevant for the energy density field in \RNAdS{d+1} in \cref{sec:sRNpertpars}. A detailed analysis of singularity structure of the differential equations, which feeds into our summary in~\cref{sec:bhGreen} can be found in~\cref{sec:bhwave}. For clarity, here we ascribe names to various singular points that occur, delineating features that are generic from those that aren't (the latter occur at higher codimension in parameter space). 
\cref{sec:Jfns} contains some details relating to the evaluation of three-point functions in~\cref{sec:example}.
Finally,~\cref{sec:localapparent} gives a quick and simple-minded discussion of deducing when a particular singular point of a second order ODE is apparent by a local Frobenius analysis (the general discussion dates back to~\cite{Heun:1888log}).

%~~~~~~~~~~~~~~~~~~~~~~~~~~~~~~~~~~~~~~~~~~~~~~~
\section{Preliminaries}
\label{sec:prelim}
%~~~~~~~~~~~~~~~~~~~~~~~~~~~~~~~~~~~~~~~~~~~~~~

Consider a general static, translational symmetric black hole  geometry with the line element in ingoing coordinates given as 
\begin{equation}\label{eq:dssq}
ds^2 = 2\, dv\, dr - r^2\, f(r)\, dv^2 + r^2 \, d\vb{x}^2\,.
\end{equation}  
We restrict attention to non-degenerate horizons with $f(r)$ having a simple zero. The largest real root $f(r)$ is taken to be $r=r_+$ (the outer horizon)  and one defines the Hawking temperature as
\begin{equation}\label{eq:HawkingT}
T = \frac{r_+^2\, f'(r_+)}{4\pi}\,.
\end{equation}  
%

%~~~~~~~~~~~~~~~~~~~~~~~~~~~~~~~~~~~~~~~~~~~~~~~
\subsection{The grSK geometry}\label{sec:grskreview}
%~~~~~~~~~~~~~~~~~~~~~~~~~~~~~~~~~~~~~~~~~~~~~~

The grSK geometry (also referred to as the grSK contour) introduced in~\cite{Glorioso:2018mmw} is a particular complexification of the black hole spacetime~\eqref{eq:dssq}, which is conjectured to be dual to the Schwinger-Keldysh contour of a thermal holographic CFT. In thermal field theory, one computes real-time observables by complexifying the time coordinate, to include the forward and backward evolution of the thermal density matrix $\rho_\beta \mapsto U(J_\skR)\, \rho_\beta\, U^\dagger(J_\skL)$. Holographically, this is achieved by complexifying the radial direction in the ingoing chart specified above.

Given the boundary Schwinger-Keldysh contour, we have a spacetime with two asymptotic boundaries, labeled L and R, above.  We will work in the modified advanced retarded (FP) basis introduced in~\cite{Chaudhuri:2018ymp}, where the sources $\JF$ and $\JP$ are the following linear combinations of the L and R Schwinger-Keldysh sources (likewise for operators)
\begin{equation}\label{eq:JFPdef}
\JF  = -\left((1+\nB)\, J_\skR - \nB\, J_\skL\right) \,, \qquad \JP = -\nB\, \left(J_\skR-J_\skL \right) \,.
\end{equation}  
Here $\nB$ is the Bose-Einstein factor $\nB = \frac{1}{e^{\beta\omega}-1}$, which we note has poles at Matsubara frequencies $\mathbf{w} = i\, m$, $m\in \mathbb{Z}$. The advantage of this basis is that the Schwinger-Keldysh, and KMS conditions are cleanly implemented as the statement of vanishing of all correlators with purely F or P insertions. Any prescription for the bulk dual, therefore, must respect these constraints. Our goal here is to argue that the grSK geometry indeed achieves this in a self-consistent manner. 
 
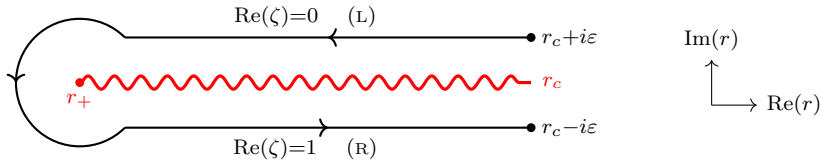
\begin{figure}[h!]
\begin{center}
\begin{tikzpicture}[scale=0.6]
\draw[thick,color=red,fill=red] (-5,0) circle (0.45ex);
\draw[thick,color=black,fill=black] (5,1) circle (0.45ex);
\draw[thick,color=black,fill=black] (5,-1) circle (0.45ex);
\draw[very thick,snake it, color=red] (-5,0) node [below] {$\scriptstyle{r_+}$} -- (5,0) node [right] {$\scriptstyle{r_c}$};
\draw[thick,color=black, ->-] (5,1)  node [right] {$\scriptstyle{r_c+i\varepsilon}$} -- (0,1) node [above] {$\scriptstyle{\Re(\ctor) =0}\quad (\skL) $} -- (-4,1);
\draw[thick,color=black,->-] (-4,-1) -- (0,-1) node [below] {$\scriptstyle{\Re(\ctor) =1}\quad (\skR) $} -- (5,-1) node [right] {$\scriptstyle{r_c-i\varepsilon}$};
\draw[thick,color=black,->-] (-4,1) arc (45:315:1.414);
\draw[thin, color=black,  ->] (9,-0.5) -- (9,0.5) node [above] {$\scriptstyle{\Im(r)}$};
\draw[thin, color=black,  ->] (9,-0.5) -- (10,-0.5) node [right] {$\scriptstyle{\Re(r)}$};  
% \draw[thick,color=blue,fill=red] (-10,0) circle (0.45ex);
% \draw[thick,color=red,fill=red] (-8,0) circle (0.45ex) node[below] {$\RQ$};
% \draw[thick,snake it, color=blue] (-10,0) node [below] {$\scriptstyle{r_-}$} -- (-15,0) ;
\end{tikzpicture}
\caption{ The complex $r$ plane with the locations of the two regulated boundaries (with cut-off $r_c$) and the outer horizon at $r_+$ marked. The grSK contour is a codimension-1 surface in this plane (drawn at fixed $v$). The direction of the contour is taken to be counter-clockwise, encircling the branch point at the outer horizon. The cut is chosen to run out towards the boundary. Depending on the system we study, there will be other branch points, cuts, and singularities in the complex radial plane. We will discuss these in the sequel, in a case-by-case basis.}
\label{fig:mockt}
\end{center}
\end{figure}

To describe the geometry, consider the mock tortoise coordinate defined as 
\begin{equation}\label{eq:mockt}
\dv{\ctor}{r} = \frac{2}{i\beta}\, \frac{1}{r^2f} \,.
\end{equation}  
This coordinate $\ctor(r)$ has a logarithmic branch cut emanating from the zeros of $f$. We will be interested in the cut running from the outer horizon, which we ruin along the ray $(r_+,\infty)$. The normalization has been chosen so that $\ctor$ has  unit monodromy across the cut. We take $\Re(\ctor) =0$ on the top leg of the grSK contour and $\Re(\ctor)=1$ on the bottom leg, as indicated in~\cref{fig:mockt}.

The basic data one needs to compute boundary Schwinger-Keldysh observables, is the boundary-bulk Green's function for the field dual to the QFT operator. Of primary interest is the one regular at the future horizon, the ingoing Green's function $\Gin{}(\ctor,v,\bx)$.  For planar black holes we exploit the Killing symmetries to work in Fourier domain $(\omega,\vb{k})$, with wavefunctions $e^{-i\,\omega\,v + i\, \vb{k}\cdot\vb{x}}$ and let $k = \abs{\vb{k}}$. We also work with a basis of the tangent space adapted to the ingoing coordinates~\cite{Ghosh:2020lel}. In particular, the derivation
\begin{equation}\label{eq:Dzplus}
\Dz_+ = r^2f\, \partial_r + \partial_v\,,
\end{equation}	
is useful to analyze  time-reversal properties.\footnote{
 All the equations we analyze are time-reversal invariant, even though the fields themselves can have intrinsic time-reversal parity. This  is important for the grSK prescription to respect the Schwinger-Keldysh and KMS conditions~\cite{Jana:2020vyx}.}  
Depending on the circumstance, it will also be useful to consider the frequencies measured in Matsubara units, so we introduce

\begin{equation}\label{eq:dimwq}
\bwt = \frac{\omega}{r_+} \,, \qquad \bqt  = \frac{k}{r_+} \,, \qquad 
\mathbf{w} \equiv \frac{\beta\,\omega}{2\pi}   \,.
\end{equation}  

The ingoing Green's function $\Gin{}(\ctor,\omega,\bk)$ satisfies suitable non-normalizable boundary conditions at the spacetime boundary, and is regular at the horizon. This is the analytic solution.  From it one can recover the outgoing boundary-bulk Green's function by using the time-reversal property of the background~\cite{Jana:2020vyx}; specifically, $\Gout{}(\ctor,\omega,\bk) = e^{-\beta\omega\ctor}\, \Gin{}(\ctor,-\omega,\bk)$. Furthermore, the bulk-bulk Green's function may also be directly obtained from the knowledge of $\Gin{}$~\cite{Loganayagam:2022zmq}.

The computation of a thermal correlation function of the boundary field theory, therefore, involves convolving these boundary-bulk and bulk-bulk Green's functions, and carrying out the radial integrals. The main difference from the usual Witten diagram computation in Euclidean-\AdS{} is that now the radial integral is carried out over the grSK geometry, using the contour depicted in~\cref{fig:mockt}. While the ingoing boundary-bulk Green's function being regular is insensitive to the branch cut, the outgoing boundary-bulk and bulk-bulk Green's functions pick up monodromies from the cut. The contour integral can, in the absence of any other singularities, be evaluated as an ordinary radial integral restricted to the domain $[r_+,\infty)$, with the integrand being given as a discontinuity of the convolved Green's functions, vertex factors, and radial Boltzmann weights (the $e^{-\beta\omega\ctor}$ factors)~\cite{Loganayagam:2022zmq}. 

%~~~~~~~~~~~~~~~~~~~~~~~~~~~~~~~~~~~~~~~~~~~~~~~
\subsection{General properties of ODEs}
\label{sec:odes}
%~~~~~~~~~~~~~~~~~~~~~~~~~~~~~~~~~~~~~~~~~~~~~~

As reviewed above, the essential ingredient in the  computation of thermal real-time correlators is the ingoing boundary-bulk Green's function. These Green's functions in Fourier domain satisfy second order ordinary differential equations.  The class of equations we are interested in are of the following form:
\begin{equation}\label{eq:homeqn}
\mathfrak{D} \sen{} + V(r)\, \sen{}= 0\,, \qquad \mathfrak{D} \equiv -\frac{1}{\sqrt{-g}\, e^\chi} \partial_A\left(e^{\chi}\, \sqrt{-g}\, g^{AB}\, \partial_B\right) .
\end{equation}  
The `dilaton' factor $e^\chi$ is a function of the radial coordinate. In certain cases it also depends on the spatial geometry, which will prove to be the ones of interest for us. The potential functions $V(r)$ can be complicated, but these will be mostly irrelevant for our local analysis.  We have compiled the various equations of interest in~\cref{sec:eqcompilation}.

Our aim is to analyze the singularity structure of the differential equations~\eqref{eq:homeqn}, which are second order, linear ODEs (SOLDE), in Fourier domain. These can be  canonically be put in the form:
\begin{equation}\label{eq:SOLDE}
\varphi''(r) + p(r) \, \varphi'(r) + q(r)\, \varphi(r) =0\,.
\end{equation}  
We let $r \in \mathbb{P}^1$ and wish to classify the branch points of the solutions $\varphi(r)$ of~\eqref{eq:SOLDE}. The interplay of this  branch structure with the contour prescription of~\cite{Glorioso:2018mmw} described in~\cref{sec:grskreview} will be important for purposes of understanding higher-point thermal correlation functions. It may be possible to reduce the number of such branch points, by projecting down from the $\mathbb{P}^1$ parameterized by $r$, say with the variable $r^2$ for even boundary dimensions. However, since we wish to work with the grSK geometry, which complexifies $r$, we will directly analyze the ODEs in this radial variable with no redefinitions.

We recall some elementary facts. A point $r = r_0$ (including the point at infinity) is an \emph{ordinary point} if $p(r)$ and $q(r)$ are analytic in an open region containing $r_0$. If not, it can be a 
\emph{regular singular point} if $(r-r_0)\, p(r)$ and $(r-r_0)^2 \,q(r)$ are analytic in an open domain of $r_0$, or an \emph{irregular singular} point otherwise. A SOLDE is Fuchsian if all of its singular points (abbreviated SPs) are  regular (modulo a monodromy condition, which will describe below). For reasons that are not a-priori clear, all the wave equations we study for perturbations of non-extremal planar \AdS{d+1} black holes are Fuchsian with regular singular points. For equations of massive scalars in such backgrounds this was already noted in~\cite{Aminov:2020yma}. 

In the neighbourhood of a regular SP $x=x_0$,  the local behaviour of the solution can be understood by the Frobenius expansion. Laurent expanding the functions $p(r)$ and $q(r)$ about $r_0$, and letting $
\{p_n,q_n\}$ be the coefficient of $(r-r_0)^n$, we learn that characteristic exponents $\alpha_\pm$ are solutions to the indicial equation, 
\begin{equation}\label{eq:indicialgen}
\alpha(\alpha-1) + \alpha\, p_{-1} + q_{-2} = 0 \,.
\end{equation}  
These tell us that the solution $\varphi(r)$ has a branch point at $r_0$ and can be used to set up a Frobenius expansion for $\varphi(r)$ in an open  neighbourhood.  This local solution picks up a monodromy upon encircling $r_0$, $r \mapsto r_0 e^{2\pi i}$, rotating by 
\begin{equation}
\begin{split}
\mqty(e^{2\pi i\, \alpha_-} & 0 \\ 0 & e^{2\pi i\, \alpha_+ }) \;\;\text{for}\;\; \alpha_+ -\alpha_- \notin \mathbb{Z}\qquad  \text{or} \qquad 
e^{2\pi i\, \alpha_-} \mqty( 1 & 2\pi i\,\gamma \\ 0 & 1)\;\;\text{for}\;\;
 \alpha_+ -\alpha_- \in \mathbb{Z}\,.
\end{split}
\end{equation}  
In the former case the two local Frobenius solutions $\varphi_\pm$ are linearly independent, while in the latter case there is a logarithmic branch.\footnote{ 
    Assuming $\alpha_+> \alpha_-$, the linearly independent solutions are  $\varphi_+$ and $\varphi_-+ \gamma\,\varphi_+ \, \log (r-r_0)  $.} 
There is however one exceptional case: for $\alpha_\pm \in \mathbb{Z}$ and $\alpha_+\neq \alpha_-$, it may be possible for there to be no monodromy -- the logarithmic branch is absent. Furthermore, if both exponents are non-negative then the solution admits a regular Taylor expansion. In such a situation the putative SP $r=r_0$ is said to be an \emph{apparent singular point} (abbreviated ASP), for it secretly is simply an ordinary point of the equation. An analysis of such ASPs, including a general criterion for their existence, can be found in~\cite{Heun:1888log}, where they are referred to as pseudo-singular points in translation.

For a Fuchsian SOLDE, the so-called Fuchs condition, requires that sum of the characteristic exponents from all the SPs, regular and apparent, (including infinity, whose exponents counted as usual with an opposite sign) is fixed in terms of the number of SP: if we have an equation with $m$ SPs, then $\sum_{i=1}^m \alpha_{i+}+ \alpha_{i-} = m-2$. A classic reference on this material is~\cite{Coddington:1955ode}; for modern perspective on Fuchsian ODEs see~\cite{Haraoka:2015ode}. 

The singularity analysis of equations compiled in~\cref{sec:eqcompilation}, which are of the form~\eqref{eq:homeqn}, is straightforward. Our interest will be in ascertaining how the singular points morph as we change the parameters in the equation. Before we do so, however, let us take stock of why this is relevant for the interaction vertices. Following this, we will summarize the basic features of the equations relevant for computing conserved current correlators in~\cref{sec:bhwave}. 

%~~~~~~~~~~~~~~~~~~~~~~~~~~~~~~~~~~~~~~~~~~~~~~~
\subsection{Bulk interactions}
\label{sec:bulkvertex}
%~~~~~~~~~~~~~~~~~~~~~~~~~~~~~~~~~~~~~~~~~~~~~~ 

We will focus on two types of bulk interactions, which require attention for checking the consistency of the grSK contour prescription. The first of these, involves interactions comprising of derivatives of fields. For simplicity,  consider the following cubic interaction of two scalar fields $\phi$ and $\chi$
\begin{equation}\label{eq:cubicFFC}
S_3 =
    \int\,d^{d+1}x\, \sqrt{-g}\ \nabla_A \, \phi\, \nabla^A \phi \,\chi\,.
\end{equation}
While it looks innocuous, recall that the choice made for the grSK contour requires the ingoing solution to be analytic, putting all the monodromies into the outgoing solution. Adapting to the ingoing tangent basis, the vertex can be written as\footnote{
    The integral with subscript $k$ is a shorthand for the momentum space integrals with the usual momentum conserving $\delta$-functions. The $d$-vector $k^\mu$ has components $k^\mu = (\omega,\vb{k})$, with $k \equiv\abs{\vb{k}}$. The plane wave basis in $\mathbb{R}^{d-1,1}$ is taken to be $e^{-i\omega v+i \vb{k}\cdot\vb{x}}$. We have also made clear that the radial integral is to be evaluated on the grSK contour~\cref{fig:mockt}. }
\begin{equation}\label{eq:cubicdervs}
S_3 =
    \int_k \,\oint\, dr\, r^{d-1}  \left[\frac{(\Dz_+ -i\omega_1 )\phi(k^\mu_1)\, (\Dz_+ + i\omega_2)\phi(k^\mu_2)}{r^2f} +\frac{\vb{k}_1\cdot \vb{k}_2}{r^2} \phi(k^\mu_1)\ \phi(k^\mu_2)\right]  \chi(k_3^\mu)\,,  
\end{equation}
where we introduced $\Dz_\pm = r^2f\, \partial_v \pm i \omega$, and have  restricted attention to translationally invariant backgrounds. The radial kinetic term here has now a simple pole at the horizon, which needs to be dealt with.  This behaviour is generic: for derivative interactions one always encounters a pole at the horizon, the order being related to the number of derivatives involved. This phenomenon was in fact noticed before in the analysis of fluctuations of a probe Nambu-Goto string~\cite{Chakrabarty:2019aeu}, where the quartic coupling of the transverse string modes have a double pole at the horizon.

The second vertex of interest is the interaction between a primary $\mathcal{O}$ of dimension $\Delta$, and the energy density operator.  The cubic vertex arises already from the kinetic terms for the scalar field
\begin{equation}\label{eq:phikinetic}
S_{_\text{EOO}} = - \int d^{d+1}x\, \sqrt{-g} \left[\frac{1}{2}\, g^{AB}\, \nabla_A \varphi \nabla_B \varphi + \frac{\Delta(\Delta-d)}{2} \, \varphi^2\right] . 
\end{equation}  
For the scalar graviton polarizations, which capture the dual of the energy density operator, we can express the metric perturbation in a suitable gauge as~\cite{He:2022jnc}:
\begin{equation}\label{eq:scalarpert}
\begin{split}
ds^2 
&= 
    ds_0^2 + \frac{\PHE - r\,f \PHW}{r^{d-3}}\, dv^2 
    + \frac{2 \left(\PHO - \PHE+r\, f\, \PHW\right)}{r^{d-1} f}  dv dr \\
& \qquad    
    - \frac{2(\PHO-\PHE) + (d-1)\, r f\, \PHW}{r^{d-1} f^2}\, dr^2 + r^2\, \frac{\PHW}{r^{d-2}}\, d\vb{x}^2\,,
\end{split}
\end{equation}  
From here it is straightforward to read off the  couplings 
\begin{equation}\label{eq:cubicEOW}
\begin{split}
S_{_\text{EOO}} 
&=
 \int d^{d+1}x  \, L_{_\text{EOO}} \\
L_{_\text{EOO}}
&= 
    -\frac{\PHE}{2\,r^2 f^2}\,\left[ (\partial_v \phi)^2 +  (\Dz_+\phi)^2 \right] 
 + \frac{\PHO}{r^2 f^2}\, \partial_v \phi \, \Dz_+ \phi 
   \frac{\PHW}{2\,rf}
 \left[ (d-1)\, (\Dz_+ \phi)^2 + m^2\, r^2 f\,  \phi^2\right] . 
\end{split}
\end{equation}  

The complication lies in the translation to the fields which satisfy simple ODEs since, owing to the background gauge invariance, the fields $\PHE, \PHO, \PHW$ are not independent degrees of freedom. As explained in~\cite{He:2022jnc,He:2022deg}, the information about the perturbation can be repackaged into gauge invariant variables (following the original analysis of ~\cite{Kodama:2003jz,Kodama:2003kk}). For a planar \SAdS{d+1} black hole, we have a single energy density operator, dual to a field $\MZ$, which determines the metric functions as~\cite{He:2022jnc}:
\begin{equation}\label{eq:EOWZ}
\begin{split}
\PHE 
&= 
    \Dz_+ \left(\frac{r}{\Lk} \left[\Dz_+ - \frac{r^2f'}{2}\right] \MZ\right) , \\
\PHO 
&= 
    \pdv{v}(\frac{r}{\Lk} \left[\Dz_+ - \frac{r^2f'}{2}\right] \MZ) , \\
\PHW 
&= 
    \frac{1}{\Lk} \left[\Dz_+ +\frac{k^2}{d-1}\right] \MZ\,.
\end{split}
\end{equation}
The function $\Lk$ which will play an important role below is defined as 
\begin{equation}\label{eq:Lkdef}
\Lk = k^2 + \frac{d-1}{2}\, r^3\, f'\,.
\end{equation}  
The corresponding expressions for the planar \RNAdS{d+1} black hole are given in \cref{sec:sRNpertpars}, since in this case we have to also deal with the mixing between the energy density and charge diffusion modes. Despite the additional complications, one finds the translation to the gauge invariant variables to involve the function $\Lk$ as in~\eqref{eq:Lkdef}, with the only change being in the metric function $f(r)$.   

For the neutral black hole, we can therefore write the vertex as
\begin{equation}\label{eq:EOOvertex}
\begin{split}
L_{_\text{EOO}}
&=
    \frac{1}{2\,r^2\,f^2}\left[
 -\Dz_+ \left(\frac{r}{\Lk} \left[\Dz_+ - \frac{r^2f'}{2}\right] \MZ\right)
 + \frac{(d-1)\, rf}{\Lk}\left[\Dz_+ +\frac{k^2}{d-1}\right] \MZ
\right] (\Dz_+ \phi)^2 
\\
&\quad 
    - \frac{1}{2\,r^2f^2}
    \Dz_+ \left(\frac{r}{\Lk} \left[\Dz_+ - \frac{r^2f'}{2}\right] \MZ\right) (\partial_v \phi)^2
    +  \frac{1}{r^2f^2} \pdv{v}(\frac{r}{\Lk} \left[\Dz_+ - \frac{r^2f'}{2}\right] \MZ)\partial_v \phi\, \Dz_+ \phi\,,
\end{split}
\end{equation}
where now the fields $\phi$ and $\MZ$ satisfy equations of the form~\eqref{eq:homeqn}. The former obeys~\eqref{eq:Mdesign} with $\ann = d-1$, while the latter obeys~\eqref{eq:Zsound}. 
 
The point to notice here is the presence of the function $\Lk$ in the denominator, which potentially can give rise to singular contributions.  We are interested in ascertaining whether its zeros are relevant, and if so whether one also has to account for the monodromy of the field $\MZ$ about these singularities.  This requires us to undertake a careful analysis of the differential equations, thus motivating the study in this paper. We note here that the factors of $\Lk$ in the vertex are due to the redefinition of the metric variables in terms of $\MZ$ in~\eqref{eq:EOWZ}. Independent of applications to real-time correlators, one would also like to know if metric perturbations are regular themselves. On physical grounds we expect them to be so for real frequencies and momenta, but the complete analysis, as it will transpire, should also allow for complexified spatial momenta. 

%~~~~~~~~~~~~~~~~~~~~~~~~~~~~~~~~~~~~~~~~~~~~~~~
\section{Green's functions on black hole backgrounds}
\label{sec:bhGreen}
%~~~~~~~~~~~~~~~~~~~~~~~~~~~~~~~~~~~~~~~~~~~~~~ 

We turn to an analysis of the wave equations encountered in the study of black hole perturbations. We will focus on the set of equations that captures holographically the dynamics of spinless operators of arbitrary conformal dimension and conserved currents, for the most part. For systems with multiple conserved currents (as in the case of finite density), diagonalize the kinetic terms for the corresponding bulk fields.  In all cases of interest, the dynamics of conserved currents can be repackaged into equations for designer scalars with non-minimal couplings. The details of the analysis (which is pretty standard) are compiled in~\cref{sec:bhwave}. We simply summarize below the features which play a role in the computation of Witten diagrams using the grSK contour prescription. 

%~~~~~~~~~~~~~~~~~~~~~~~~~~~~~~~~~~~~~~~~~~~~~~~
\subsection{Schwarzschild-AdS black hole}\label{sec:sads}
%~~~~~~~~~~~~~~~~~~~~~~~~~~~~~~~~~~~~~~~~~~~~~~

Let us begin with a class of equations that encompasses minimally coupled massive scalar fields dual to CFT primaries, probe gauge fields dual to a global current operator, and tensor and vector polarizations of gravitons. As demonstrated in~\cite{Ghosh:2020lel} all these cases are described by the dynamics of a designer scalar field ~\eqref{eq:homeqn} with a simple power law behaviour of the auxiliary dilaton, $e^\chi \sim r^{\ann-d+1}$. The equation of interest~\eqref{eq:Mdesign} translates to 
\begin{equation}\label{eq:SLMdesign}
\varphi''(r) 
+ \left[\frac{\ann+2}{r} + \frac{f'}{f} - \frac{2i\omega}{r^2f}\right] \varphi' 
- \left[\frac{k^2 + m^2 \, r^2  + i\, \ann \,\omega\,r }{r^4\,f}\right] \varphi =0\,.
\end{equation}  

\begin{figure}[t!]
\centering
\begin{tikzpicture}[scale=1.0]
\draw[very thin,color=black,dashed] (-5,0) circle (2cm);
\draw[thick,color=red,fill=red] (-5,0) circle (0.25ex) node[below] {$0$};
\draw[thick,color=red,fill=red] (-7,0) circle (0.25ex) node[left] {$-r_+$};
\draw[thick,color=red,fill=red] (-3,0) circle (0.25ex) node[left] {$r_+$};
\draw[thick,color=red,fill=red] (-5,2) circle (0.25ex) node[above] {$i r_+$};
\draw[thick,color=red,fill=red] (-5,-2) circle (0.25ex) node[below] {$-i r_+$};
\draw[thick,color=red,fill=red] (1,0) circle (0.25ex) node[below right] {$\infty$};
\draw[thick,snake it, color=blue] (-3,0) -- (1,0) ;
\draw[thick,color=black, ->-] (-3,-0.3) -- (1,-0.3);
\draw[thick,color=black,->-] (1,0.3) -- (-3,0.3);
\draw[thick,color=black,-<-] (-3,-0.3) arc (315:45:0.424);
\end{tikzpicture}
\caption{ Singular points  of \SAdS{5} designer scalar equation in the complex $r$. We have drawn a circle of radius $r_+$ that separates the interior of the black hole from the exterior. The cut depicted is the one running from the horizon to the \AdS{5} boundary for the outgoing Hawking mode (we make the choice that the ingoing mode is analytic). The grSK contour encircles this cut counter-clockwise as shown. The cuts from other singular points are not indicated, but can run in any directions so long as they do not cross the grSK contour.}
\label{fig:designerSchw}
\end{figure}
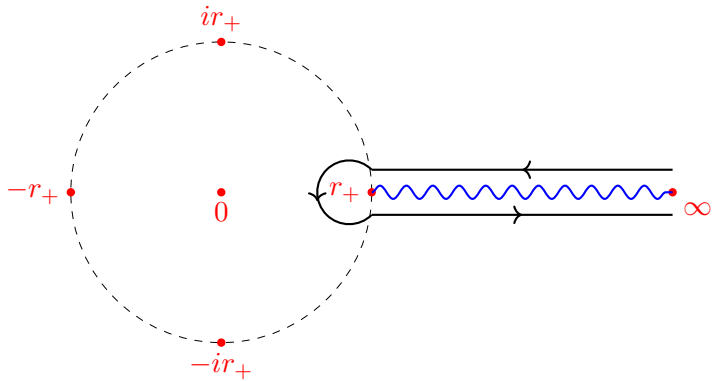

This system has SPs  at the asymptotic boundary, at the black hole curvature singularity (the origin), and at the zeros of $f$ for generic values of $\omega,k$.  We have depicted this generic behaviour in~\cref{fig:designerSchw} (for $d=4$). 

The behaviour at infinity dictates the familiar asymptotic fall-offs. The only other SP relevant for the grSK contour, is that originating from the outer horizon $r=r_+$, where the characteristic exponents are $0$ and $i\frac{\beta\omega}{2\pi}$, corresponding to the ingoing and  outgoing solution, respectively.
Using the freedom to orient cuts from the other SPs away from the grSK contour, we see that there is no ambiguity in the prescription. Thus, in this case we only need to deal with issues arising from bulk vertices, which as noted in~\cref{sec:bulkvertex} may themselves contribute singularities. 

There are additional features at special values of frequency and momenta. At Matsubara frequencies $\bwt = i\, n$ with $n\in \mathbb{Z}_{\geq0}$ the outgoing solution can be made regular at the horizon by fine-tuning the momentum. For this choice the horizon is an ASP. While we explore these in~\cref{sec:schw}, in the context of grSK contour, their presence simply implies that the correlators are analytic at these special points.

There is one equation~\eqref{eq:Zsound}, that for the bulk mode $\MZ(r)$ dual to the energy density operator, which has to analyzed independently. Naively, it appears that in addition to the asymptotic, curvature, and horizon SP, we have an additional $d-2$ SPs from the roots of $\Lk=0$. For \SAdS{d+1} are located on circle of size $1/k$ at 
\begin{equation}\label{eq:rkschw}
r_k = r_+ \, \left(-\frac{2}{d\,(d-1)}\, \frac{1}{\bqt^2}\right)^\frac{1}{d-2} \, \varpi_{d-2}\,,
\end{equation}  
where $\varpi_n$ denotes the $n^{\rm{th}}$ roots of unity. For $\arg(\bqt^2) =\pi$ we even potentially have one of the roots located between the horizon and the asymptotic boundary. However, a closer analysis reveals that for generic $\bqt^2$ the points $r_k$ are apparent SPs (ASPs) of~\eqref{eq:Zsound}.  

At special values of parameters, we find new behaviour. The roots of $\Lk$ can be made to coalesce with other SPs of the system, by tuning the momentum. This leaves the Fuchsian nature of the equation unchanged, but modifies the characteristic exponents. There are two cases of interest: $r_k \to \infty$ and $r_k \to r_+$. 

When $k=0$, the roots of $\Lk$ are at infinity. However, this leaves the point at infinity a regular SP, albeit with changed exponents. As was understood in~\cite{He:2022jnc} the change in behaviour owes to enhanced diffeomorphism symmetry in the zero momentum limit (a fact already appreciated in~\cite{Kodama:2003jz} and explained in part in~\cite{deOliveira:2018jhc}).\footnote{
    In fact, there are additional zero modes not visible to the bulk equation of motion, as they are purely localized on the boundary~\cite{He:2022jnc}.}

By choosing 
\begin{equation}
k = \pm i \, \sqrt{2\pi \,(d-1)\,\frac{T}{r_+}}\,,
\end{equation}  
we can achieve $r_k = r_+$. For generic $\omega$,  the only change is that the characteristic exponents are shifted by unity: the ingoing mode has exponent $+1$, and the outgoing mode has exponent $1+i \frac{\beta\omega}{2\pi}$. 
At a particular fine-tuned values of $(\omega ,k)$, setting $\omega = 2\pi i\, T$ the energy horizon coalescence renders the horizon to be an ASP. At this point profile of $\MZ$ along the boundary directions takes the form:
\begin{equation}\label{eq:Lywave}
\exp\left(2\pi\,T \left(v- \frac{\abs{\vb{x}}}{v_B} \right) \right) \,, \qquad v_B = \sqrt{\frac{d}{2\,(d-1)}}\,.
\end{equation}  
The result is suggestive of an exponentially growing mode in time at a rate set by the maximal Lyapunov exponent $\lambda_L = 2 \pi T$ with the spatial dependence being damped at a rate given by the butterfly velocity. These are in fact, the values computed for \SAdS{d+1} black holes using the shockwave approximation~\cite{Shenker:2013pqa} (see also~\cite{Roberts:2014isa}). As noted earlier, this phenomenon was first noticed in~\cite{Grozdanov:2017ajz} and explained carefully in~\cite{Blake:2018leo}. In the SYK model it is the presence of such an exponentially growing mode that leads to the growth of the out-of-time-order correlation functions~\cite{Maldacena:2016hyu}. Inspired by this, an effective theory of maximal chaos was proposed in~\cite{Blake:2017ris}, arguing for there to be a growing mode in the energy density correlator. 

We have simply verified these earlier observations (trivially extending them for all $d\geq 3$). At this particular value of parameters, as with the Matsubara ASP, there is an ambiguity in the definition of the boundary retarded Green's function. In the present case this is just as well, since an exponentially growing mode would have signaled an instability (the quasinormal spectrum is required to be supported in the lower half-plane with retarded Green's functions analytic above the real frequency axis). We are also wary of referring to this phenomenon as `pole-skipping' for similar reasons, and prefer the term, apparent anti-quasinormal mode.

In summary, the singularity structure for all the ODEs of interest in \SAdS{d+1} is captured by the designer scalar equation~\eqref{eq:SLMdesign}, as depicted in \cref{fig:designerSchw} for $d=4$. The regular SPs are only at the roots of $f$, the boundary and at $r=0$. The nature of these regular singular points changes due to coalescence as we tune $k$ to some special values, but apart from that there is no other singularity to deal with. 

%~~~~~~~~~~~~~~~~~~~~~~~~~~~~~~~~~~~~~~~~~~~~~~~
\subsection{Reissner-Nordstr\"om-AdS black hole}\label{sec:rnads}
%~~~~~~~~~~~~~~~~~~~~~~~~~~~~~~~~~~~~~~~~~~~~~~

\begin{figure}[ht!]
\centering
\begin{tikzpicture}[scale=1]
\draw[very thin,color=black,dashed] (-5,0) circle (2cm);
\draw[very thin,color=black,dotted] (-5,0) circle (0.54858cm);
\draw[very thin,color=black,dotted] (-5,0) circle (1cm);
\draw[thick,color=gray,fill=gray] (-5,0) circle (0.25ex) node[below] {$\scriptstyle{0}$};
\draw[thick,color=red,fill=red] (-7,0) circle (0.25ex) node[left] {$\scriptstyle{-r_+}$};
\draw[thick,color=red,fill=red] (-3,0) circle (0.25ex) node[left] {$\scriptstyle{r_+}$};
\draw[thick,color=red,fill=red] (-5.54858,0) circle (0.25ex) node[above] {$\scriptstyle{-r_-}$};
\draw[thick,color=red,fill=red] (-4.51413,0) circle (0.25ex) node[above] {$\scriptstyle{r_-}$};
\draw[thick,color=red,fill=red] (-5,2.058) circle (0.25ex) node[above] {$\scriptstyle{i r_i}$};
\draw[thick,color=red,fill=red] (-5,-2.058) circle (0.25ex) node[below] {$\scriptstyle{-i r_i}$};
\draw[thick,color=PineGreen,fill=PineGreen] (-4,0) circle (0.25ex) node[below] {$\scriptstyle{\RQ}$};
\draw[thick,color=PineGreen,fill=PineGreen] (-6,0) circle (0.25ex) node[below] {$\scriptstyle{-\RQ}$};
\draw[thick,color=cyan,fill=cyan] (-3.2,2) circle (0.25ex) node[below] {$\scriptstyle{r_{k2}}$};
\draw[thick,color=cyan,fill=cyan] (-3,-2) circle (0.25ex) node[below] {$\scriptstyle{\bar{r}_{k2}}$};
\draw[thick,color=red,fill=red] (1,0) circle (0.25ex) node[below right] {$\scriptstyle{\infty}$};
\draw[thick,snake it, color=blue] (-3,0) -- (1,0) ;
\draw[thick,color=black, ->-] (-3,-0.3) -- (1,-0.3);
\draw[thick,color=black,->-] (1,0.3) -- (-3,0.3);
\draw[thick,color=black,-<-] (-3,-0.3) arc (315:45:0.424);
\end{tikzpicture}
\caption{ Singular points  encountered on complex radial plane for the various differential equations of interest in the \RNAdS{5} background.  The singular points $\{\pm r_+, \pm r_-, \pm i\,r_i $  are the roots of $f(r)$, which along with the point at infinity are SPs of all the equations of interest.  The black hole singularity is not a SP in all cases, but is for some of the equations. Additionally, we have some equations (transverse vector perturbations of the gauge field) for which the roots of $f'(r)$ located at $\pm \RQ$ are also singular. Finally, $r_{k2}$ and $\bar{r}_{k2}$ are the roots of $\Lk$ which are regular SPs of the field dual to the charge diffusion operator.  }
\label{fig:RNsummary}
\end{figure}
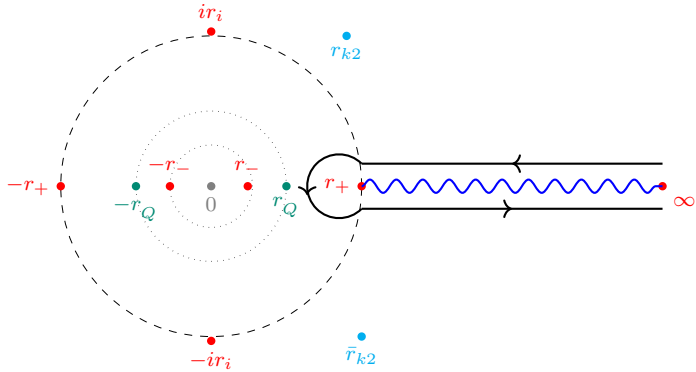

For the \RNAdS{d+1} black hole, the analysis of the differential equations is quite similar. The main differences are that depending on the equation we study there are, in addition to the SPs at infinity, the curvature singularity, and the roots of $f(r)$, additional features at specific radial positions.  One of  these corresponds to the roots of $f'(r)$, denoted $\RQ$, defined by
\begin{equation}\label{eq:RQsdcdef}
\RQ^{d-2} = \frac{d-1}{d}\, \frac{2\,Q^2}{1+Q^2}   \, r_+^{d-2}  \,, \qquad  \sdc \equiv  \frac{\RQ^{d-2}}{r_+^{d-2}} \,.
\end{equation}  
The dimensionless parameter $\sdc$ is serves as a proxy for the charge, with $\sdc =0$ in the neutral case, and $\sdc =1$ for the extremal black hole. The other is the roots of the function $\Lk$, which are distributed in two sets on circles of radii $r_{k1}$ and $r_{k2}$, respectively. Of these $r_{k1}$ is analogous to the locus $r_k$ for the \SAdS{d+1} black hole; it is an apparent singular point. The locus $r_{k2}$, however, is a genuine SP for the charge diffusion mode. There are of course some special points in the parameter space where the nature of the SPs changes (eg., at Matsubara frequencies for fine-tuned momenta at the horizon SP).  Details of these can be found in~\cref{sec:rnadswave}, but the general picture is summarized in~\cref{fig:RNsummary}.

There is one particular situation of fine-tuning that we wish to emphasize here, involving the roots of $\Lk$ lying at the horizon, and leading to apparent SPs. The perturbations involved are the field $\Zd$ dual to the energy density operator, and $\Vd$ dual to the charge diffusion operator, which are analyzed in~\cref{sec:sRNads}. These two satisfy decoupled wave equations with the function $\Lk$ controlling both their kinetic terms. The roots of $\Lk$ satisfy $r_{k2} < r_+ < r_{k1}$ for $\sdc \in [0,\frac{1}{2})$, but are ordered as $r_{k1} < r_+ < r_{k2}$ for $\sdc \in [\frac{1}{2},1)$.

When the dimensionless ratio $\sdc < \frac{1}{2}$, and $k$ chosen so that 
$r_{k1} = r_+$, we find that the horizon is an apparent singularity for the energy density field $\Zd$, at a specific frequency $\omega = 2\pi i\,T$. However, there is no choice of frequencies for which the field $\Vd$ is analytic, and so the horizon remains a regular SP for this field.

At this value of parameters, the profile of $\Zd$ is as in~\eqref{eq:Lywave}, viz., an exponentially growing mode. We recover from this profile the Lyapunov exponent for a neutral probe.\footnote{
    While it has been argued that in the presence of global conservation laws the bound on the Lyapunov exponent is modified to $\frac{2\pi T}{1-\abs{\frac{\mu}{\mu_c}}}$,~\cite{Halder:2019ric}, our probes are the conserved currents which are uncharged under the symmetry. 
} Furthermore, from the value of the momentum at which the outgoing mode remains analytic, we recover the butterfly velocity. We conclude that by fine-tuning parameters, the energy density field can have a putative exponential growing mode, which is analytic at the horizon, with the wavefunction in the boundary directions behaving as 
\begin{equation}\label{eq:RnvB}
\exp\left(2\pi T \left(v-\frac{\abs{\vb{x}}}{v_B}\right)\right) \,, \qquad 
v_B = \sqrt{\frac{d-(d-2)\,Q^2}{2\,(d-1)}}\,, \qquad 0 < \sdc < \frac{1}{2} \,.
\end{equation}  
This is indeed the expected value of the butterfly velocity for a \AdS{} black hole~\cite{Mezei:2018jco}. For this range the charge density field has a non-analytic outgoing mode at the horizon.

For $\sdc>\frac{1}{2}$, however, something very different occurs. In this range of charges we can tune $r_{k2} =r_+$ by choosing the momentum appropriately. We find that energy density mode $\Zd$  is always non-analytic at the horizon, for there is no choice of frequency that makes the outgoing mode regular. However, the horizon can be made into an apparent singularity for the charge diffusion mode, albeit at a different frequency, $\omega = -2\pi i\, T$. This frequency leads to an exponentially damped mode, that \emph{cannot} be identified as having to do with any kind of Lyapunov behaviour (and thus with the chaotic behaviour of the OTOC). 
% Details of this analysis can be found in~\cref{sec:sRNads}.

This change in behaviour of the fields the horizon occurs for $\sdc = \frac{1}{2}$, which translates to a \RNAdS{d+1} black hole having a charge parameter 
\begin{equation}
Q_* = \sqrt{\frac{d}{3\,d-4}} \,.
\end{equation}  
The parameterization of the metric in~\eqref{eq:fRN} with $0 \leq Q \leq \sqrt{\frac{d}{d-2}}$, the upper limit corresponding to the extremal solution.
When the charge of the black hole is small $Q<Q_*$, the behaviour we find for the energy density mode $\Zd$ is analogous to that for a neutral black hole, by continuity. However, there is a transition at a critical charge $Q_*$ (because there is a secondary branch of zeros). For large charges there is no exponentially growing analytic mode, in either the charge density, or in the energy density operators.  While in neither case is one actually computing the OTOC directly,  the essential point is that in the absence of a temporally growing mode. Therefore, there can be no meaningful association with chaos for large enough charge. As far as we can tell this point has been missed in earlier analyses~\cite{Abbasi:2020ykq,Jansen:2020hfd}. We will examine what these features mean in the Schwinger-Keldysh correlator, in due course.

%~~~~~~~~~~~~~~~~~~~~~~~~~~~~~~~~~~~~~~~~~~~~~~~
\section{Bulk interactions and the grSK contour}
\label{sec:intvertex}
%~~~~~~~~~~~~~~~~~~~~~~~~~~~~~~~~~~~~~~~~~~~~~~

Having ascertained the behaviour of solutions to the wave equations, we can now turn to the analysis of boundary correlation functions. As described in \cref{sec:bulkvertex}, our primary concern is with bulk vertices that potentially have singularities that interfere or pinch the grSK contour. The locations of the SPs along with the grSK contour for the equations of interest are given in \cref{fig:designerSchw,fig:RNsummary} (for specific equations in the charged case, see~\cref{fig:designerRN,fig:YRN,fig:VRN}). 
  
%~~~~~~~~~~~~~~~~~~~~~~~~~~~~~~~~~~~~~~~~~~~~~~~
\subsection{Interaction terms with horizon poles}
\label{sec:horpole}
%~~~~~~~~~~~~~~~~~~~~~~~~~~~~~~~~~~~~~~~~~~~~~~

Let us being with the cubic interaction of derivative coupled scalars~\eqref{eq:cubicdervs} where the vertex function naively has a simple pole at the horizon. To process this vertex we will need to understand which correlation function is affected by the presence of the pole.  In particular, we first need to check that the SK and KMS conditions are not spoiled by the presence of the horizon pole in the vertex. Furthermore, we need to  determine which of the non-vanishing correlators get a contribution from it and estimate them.

To proceed, we then recall some useful facts about the $\Dz_\pm$ derivatives that was used to relate the outgoing boundary-bulk Green's function to the ingoing one~\cite{Jana:2020vyx}. In terms of the ingoing adapted derivatives satisfy a conjugation property 
\begin{equation}\label{eq:Dpmconj}
\Dz_\pm = r^2f\, \partial_r \pm \partial_v \,, \qquad
e^{\beta\omega\zeta}\, \Dz_+ e^{-\beta\omega\zeta} = \Dz_-\,.
\end{equation}  
Using this we can argue that, given an ingoing boundary-bulk Green's function $\Gin{}(\ctor,k^\mu)$, the outgoing Green's function is given by   
\begin{equation}\label{eq:Ginoutconj}
\begin{split}
\Gout{}(\ctor,k^\mu) &= e^{-\beta\omega\ctor}\, \Gin{}(\ctor,\wk^\mu) \\
(\Dz_+ \mp i \omega)\Gout{}(\ctor,k^\mu)
&=
    e^{-\beta\omega\zeta}\,  (\Dz_- \mp i \omega)  \Gin{}(\zeta,\wk^\mu)\,.
\end{split}
\end{equation}  
Here $\wk^\mu$ is the frequency reversed $d$-momentum; if $k^\mu = (\omega,\vb{k})$, then $\wk^\mu = (-\omega,\vb{k})$.  In terms of this data the solution to the free wave equation for a field $\phi$ on the grSK geometry with sources $J_\skR$ and $J_\skL$ on the R and L boundaries is given  in the FP basis by (cf.,~\cite{Jana:2020vyx}) 
\begin{equation}\label{eq:}
\phi(\ctor,\omega,\vb{k}) = -\Gin{\phi}(\ctor,\omega,\vb{k}) \, \JF  + 
\Gout{\phi}(\ctor,\omega,\vb{k})\, e^{\beta\omega}\, \JP\,.
\end{equation}  

Next, we note that suitable combinations of the $\Dz_+$ derivative and temporal derivative lead to factors of $f$, viz., 
\begin{equation}
\begin{split}
(\Dz_+ + i \omega)\Gin{}(\ctor,k^\mu)  
&= 
    r^2f\,\pdv{r}\Gin{}(\ctor,k^\mu) \,, \\
(\Dz_+ - i\omega)\Gout{}(\ctor,k^\mu)  
&= 
    r^2f\,e^{-\beta\omega\ctor} \, \pdv{r}\Gin{}(\ctor,\wk^\mu) \,.    
\end{split}
\end{equation}
Hence, the only case of concern is one where there is a non-trivial term without a factor of $f$, which occurs when we have a contribution of the form
\begin{equation}
(\Dz_+ + i \omega)\Gout{}(\ctor,k^\mu)  \,.
\end{equation}  

With this information, we now go over the various 3-point functions and check what the presence of this pole does.  For the vertex arising from~\eqref{eq:cubicdervs} we will use $\Gin{\phi}(\ctor,k^\mu)$ and $\Gin{\chi}(\ctor,k^\mu)$ to denote the Green's functions for the two fields. We also order the fields with momentum labels $k_1^\mu$ and $k_2^\mu$ assigned to the two $\phi$ fields, and $k_3^\mu$ assigned to $\chi$. The arguments below are a specialization of those given in~\cite{Jana:2020vyx} (see also~\cite{Loganayagam:2022zmq}) so we will be brief in our exposition.

\begin{itemize}[wide,left=0pt]
\item  The horizon pole is benign for the FFF correlator which vanishes as desired. Here factor $(\Dz_+ +i\omega)\Gin{\phi}(\ctor,k_2)$  gives a factor of $r^2f$ which cancels the pole. The vanishing of the correlator follows along the lines then described in the aforementioned references. A similar argument applies to the FFP correlator.
\item Using a similar property of $(\Dz_+ -i\omega)\Gout{\phi}(\ctor,k_2)$ as noted above, we can check that PPP, PPF and PFF correlators (nb: last label is for $\chi$) are unaffected. In particular, PPP correlator vanishes, and the other two are computed as the integral of the discontinuity across the cut.  
\item This leaves us with FPF and FPP, which are both non-trivial since there is no factor of $f$ from the derivatives acting on the Green's function for $\phi$. The FPF correlator turns out to be computed by the integral
\begin{equation}\label{eq:S3FPF}
\begin{split}
I_{_\text{FPF}}
&= 
    \int_k \,\oint\, dr\, r^{d-1} \,e^{-\beta\omega_2\zeta}\Gin{\chi}(\zeta,k_3^\mu)\, \left[\frac{(\Dz_+-i\omega_1 )\Gin{\phi}(\zeta,k_1^\mu)\, (\Dz_- +i\omega_2)\Gin{\phi}(\zeta,\wk_2^\mu)}{r^2f} \right.\\
&\left.\hspace{6cm} 
    + \,\frac{\vb{k}_1\cdot\vb{k}_2}{r^2} \Gin{\phi}(\zeta,k_1^\mu) \Gin{\phi}(\zeta,\wk_2^\mu)\right]   ,
\end{split}
\end{equation}  
where we have made use of the relations deduced above. Now we have an explicit horizon pole in the integrand. A similar story pertains for FPP, which  requires us to look at the following integral:
\begin{equation}\label{eq:S3FPP}
\begin{split}
I_{_\text{FPP}}
&= 
    \int_k \,\oint\, dr\, r^{d-1} \,e^{-\beta(\omega_2+\omega_3^\mu)\zeta}
    \Gin{\chi}(\zeta,\wk_3)\, \left[\frac{(\Dz_+ -i\omega_1 )\Gin{\phi}(\zeta,\wk_1)\, (\Dz_- +i\omega_2)\Gin{\phi}(\zeta,\wk_2)}{r^2f} \right.\\
&\left.\hspace{7cm} 
    + \,\frac{\vb{k}_1\cdot\vb{k}_2}{r^2} \Gin{\phi}(\zeta,k_1^\mu) \Gin{\phi}(\zeta,\wk_2^\mu)\right]   .
\end{split}
\end{equation}  
In both cases the first lines of the expressions have a pole at the branch point, which needs evaluation. 
\end{itemize}

To get a sense of how to tackle these integrals with a pole on the branch point, let us consider the following toy integral (here $C_H$ is the keyhole or Hankel contour as  in \cref{fig:designerSchw})
\begin{equation}
\begin{split}
I&= 
    \oint_{C_H} \frac{dz}{z}\, z^{\alpha}\, e^{-z} 
= 
\left[\int_{\infty +i0}^{\epsilon+i0} +  \int_{\epsilon -i0}^{\infty-i0} \right]
\frac{dx}{x}\, x^{\alpha}\, e^{-x} + \int_{\epsilon+i0}^{\epsilon-i0}\,  \frac{dz}{z}\, z^{\alpha}\, e^{-z} \,,  \\
&=
    (e^{2\pi i\,\alpha}-1) \left[\int_{\epsilon}^\infty\, \frac{dx}{x}\, x^{\alpha}\, e^{-x} + \frac{\epsilon^{\alpha}}{\alpha} \right] . 
\end{split}
\end{equation}  
The instinct is to realize that brackets in the last line involve two divergent pieces, which should cancel amongst themselves. The result is then
\begin{equation}
I = (e^{2\pi i\,\alpha}-1)\, \Gamma(\alpha) \,,
\end{equation}  
which is manifestly analytic as a function of $\alpha$.

The integral we want with the horizon pole is something more involved. Moreover, the Green's functions do not have exponential damping but only power-law behaviour in the radial direction. However, we can use the above example to inspire a simple prescription for the computation. Let us therefore consider the following  integral which captures both correlators of interest (with $\text{x} \in\{\text{F},\text{P}\}$):
\begin{equation}\label{eq:polecutI}
I_{\text{FPx}} = \oint \frac{dr}{r^2f} \, e^{-\beta\omega\,\ctor}\, X(\omega,\ctor) \,.
\end{equation}  
We have isolated the dependence on the frequency that enters the monodromy and suppressed dependence on all other parameters. The function in the integrand, $X$, is obtained from the various boundary-bulk propagators, measures etc. The only information we need from it, is that it is analytic at $r=r_+$. 
By the above logic
\begin{equation}\label{eq:polecutfinal}
\begin{split}
I_{\text{FPx}} 
&= 
    (e^{-\beta\omega} -1) \, \int_{r_++\epsilon}^{\infty} \frac{dr}{r^2f} \, e^{-\beta\omega\,\ctor}\, X(\omega,\ctor) + 
    \frac{\beta}{4\pi}\, X(r_+)\, \epsilon^{\frac{i\beta\omega}{2\pi}} \,  2\pi\frac{e^{-\beta\omega}-1}{i\beta\omega} \\
&=
     (e^{-\beta\omega} -1) \left[\int_{r_++\epsilon}^{\infty} \frac{dr}{r^2f} \, e^{-\beta\omega\,\ctor}\, X(\omega,\ctor) 
     + \frac{X(r_+)}{2i\omega} \, \epsilon^{\frac{i\beta\omega}{2\pi}} 
     \right]   . 
\end{split}
\end{equation}  

The final result involves a localized contribution from the horizon the serves to cancel the IR divergent term in the radial integral, giving thereby a sensible result for the boundary correlation function. In~\cite{Chakrabarty:2019aeu}, such a localized contribution was extracted by working in a small frequency gradient expansion. 

From our perspective, the contour integral prescription of~\cite{Glorioso:2018mmw} has the advantage of making clear how to deal with these horizon poles. In the earlier prescriptions of~\cite{Skenderis:2008dg}, as for example recently employed in~\cite{Pantelidou:2022ftm}, the piecewise integrals will each have IR divergences, which will need regulating. The contour prescription provides one such, and has the added benefit of being geometric. We will employ this to compute an explicit correlator in~\cref{sec:example} below.

%~~~~~~~~~~~~~~~~~~~~~~~~~~~~~~~~~~~~~~~~~~~~~~~
\subsection{Coupling to energy density}
\label{sec:encoupling}
%~~~~~~~~~~~~~~~~~~~~~~~~~~~~~~~~~~~~~~~~~~~~~~

The other coupling of interest is the coupling to the energy density operator. We have  explicitly  written out the cubic coupling of a primary of dimension $\Delta$ to the energy density operator~\eqref{eq:EOOvertex}. There are several terms in $S_{_\text{EOO}}$, but all have a common factor of $\Lk^{-1}$. There are additionally potential horizon poles. We expect something similar for cubic self-coupling of the scalar graviton polarization with itself. For instance, there will be a $\MZ^3$ vertex (with some complicated action of derivatives), which will contain an overall $\Lk^{-3}$. 

While computing the contributions to the $\expval{T_{\mu\nu} \mathcal{O} \mathcal{O}}$ three-point function, naively, one  will have to evaluate integrals like
\begin{equation}\label{eq:Lkintegrand}
\oint \frac{dr}{r^2f\, \Lk} \, e^{-\beta\omega_2\,\ctor}\, X(\omega_1,\ctor) \,.
\end{equation}  
The horizon pole contributions can be understood as in~\cref{sec:horpole} (later we will argue that some factors present in~\eqref{eq:EOOvertex} are spurious). Let us therefore focus on the $\Lk$ contribution for now.

If the zero of $\Lk$, $r=r_k$ lies outside the contour of integration, then we do not need to concern ourselves, and the calculation proceeds as usual. The key issue is when the grSK contour is pinched by the pole at $r_k$, which, as we have seen,  can occur for complex momenta. For instance, restricting attention to the \SAdS{d+1} case, taking $\arg(\bqt^2) = \pi$  will achieve this.\footnote{
    For \RNAdS{d+1} we can attain this by requiring $\arg(\BQTs^2) =\pi$ $\sdc<\frac{1}{2}$ or $\arg(\BQTs^2) =0$ when $\sdc>\frac{1}{2}$, respectively (cf.,~\cref{sec:sRNads}).
} 
It is also worth emphasizing that if we focus on small $k$ and analyze the problem in a boundary gradient expansion, as has been done in earlier works~\cite{He:2022jnc,He:2022deg}, then the zeros of $\Lk$ do not show up.

The ingoing Green's functions for the fields dual to the energy density operator (or charge diffusion)  do not have a branch cut at this point from the SOLDE analysis. However, the explicit pole in the integrand~\eqref{eq:Lkintegrand}  cannot be avoided. This is a serious problem for the grSK contour, unless there is a mitigating effect from the rest of the integrand. This is because, should the function $X(\omega_1,\ctor)$ have a constant limit at $r_k$, then we will pick up this value, as a localized residue contribution at the pole. Such a contribution will violate the SK and KMS conditions, rendering the grSK prescription suspect. The easiest way to see this is to note that one can end with a non-vanishing answer for the correlator for all F-type operators with energy density insertion. While the usual Schwinger-Keldysh rules will imply vanishing in the boundary, $X(\omega_1,\ctor)$ will be a product of ingoing Green's function, which are analytic, and may have a non-vanishing residue (setting  $\omega_2 =0$ for simplicity). The only way to avoid this would be for the residue to vanish.  

In other words, consistency with the Schwinger-Keldysh rules, requires that the all interactions with the energy density operator, have suitable zeros in the numerator to cancel the inverse powers of $\Lk$. The origin of the $\Lk^{-1}$ factors in the integrand is from the field redefinitions~\eqref{eq:EOWZ} between the metric fields $\{\PHE,\PHO, \PHW\}$ and $\MZ$. Explicitly, carrying out a local Frobenius expansion of $\MZ$ in the neighbourhood of $r_k$, we can that the integrand has no poles for any interaction involving the energy-momentum tensor's scalar polarization. 

As such, one would find it strange for the metric functions to diverge in the physical region between the horizon and the boundary for complex momenta. Lest the reader imagine that this would be guaranteed, we should clarify that the intuition which holds for real momenta could have failed for complex momenta. That this does not happen involves non-trivial cancellations as can be checked algebraically. Once this is established, one can directly use~\eqref{eq:cubicEOW} to deduce that the correlation functions remain well-defined even for $\arg(\bqt^2) =\pi$ in the \SAdS{d+1} geometry. 

In fact, a similar reasoning can be applied to the potential horizon poles 
in~\eqref{eq:EOOvertex}. Reverting to the metric perturbation itself, say~\eqref{eq:scalarpert}, which has explicit factor of $f$ in the denominator, we recall, that the natural cotangent basis in ingoing coordinates involves the one-form $\frac{dr}{r^2f}$. One can see that all the factors of $1/f$ are part of such a one-form, which secretly guarantees regularity of the metric perturbation at the horizon. However, in the kinetic term of the field $\phi$ dual to the primary $\mathcal{O}$ the factors of $1/f$ are physical, and do contribute as in~\eqref{eq:polecutfinal}. 

For \RNAdS{d+1} we have to deal with an additional complication. The field $\Vd$ actually diverges at $r=r_{k2}$. This seems more serious;~\eqref{eq:EOWMZ} and~\eqref{eq:MZVdiagonal} suggest that there might be double-poles in the metric and gauge field perturbations. However, once again, the physical metric and gauge field perturbations are regular at the zeros of $\Lk$. This can be directly checked using the local Frobenius solution of  $\{\Zd,\Vd\}$. 

The upshot of our analysis is that the zero loci of the function $\Lk$, which was introduced to find autonomous SOLDEs for the perturbations, are irrelevant for the purposes of computing any physical observable. These zeros are for the most part ASPs of the differential equation (the exception being the one set of zeros in the charged black hole for $\Vd$), and not only are the decoupled fields regular there, but the physical metric and gauge field functions are themselves regular. The skeptical reader might inquire whether this was all a red herring from the beginning. Could one have dispensed with fields like $\MZ$ in the \SAdS{} (or $\Zd$ and $\Vd$ in the \RNAdS{} case) and worked with the metric functions directly? The function $\Lk$ was originally presented in~\cite{Kodama:2003jz,Kodama:2003kk}, and a (technical) rationale was its origin was offered in~\cite{He:2022jnc,He:2022deg}. What we can assert with some confidence is that none of the metric functions themselves satisfy an autonomous SOLDE; one just has a coupled system of first and second order equations (see the latter works for a thorough analysis of the dynamics).  While there is some satisfaction in noting that there are no special loci in the geometry determined for a subset of momenta, we do not have at present more physical picture to paint for the existence of $\Lk$ and its irrelevance in the final answers. 

%~~~~~~~~~~~~~~~~~~~~~~~~~~~~~~~~~~~~~~~~~~~~~~~
\subsection{Apparent singularities at the horizon}
\label{sec:apphor}
%~~~~~~~~~~~~~~~~~~~~~~~~~~~~~~~~~~~~~~~~~~~~~~

The preceding analysis focused on having additional poles in the integrand either at the horizon, which generically is a regular SP, or at $r_k$ which is always an ASP of the SOLDEs. Let us now turn to the case where the horizon itself becomes an ASP, at special codimension-2 loci in the complexified frequency domain. There are two situations we should consider here: the Matsubara ASP, which pertains for all the equations analyzed herein, and the case where the energy ASP at $r=r_k$ merges with the horizon leading to energy horizon coalescence described in \cref{sec:energyLk} and \cref{sec:sRNads}. In these cases, particular values of frequency and momenta render the horizon an ASP of the corresponding equation. Even in the exceptional case of charge diffusion mode $\Vd$, at the coalescence of a root from the set $r_{k2}$ with $r_+$, we encounter an ASP along a codimension-2 locus.

Consider first the Matsubara ASP, where for $\omega = -2\pi i \,m\,T$, with $m \in \mathbb{Z}_+$ and fine-tuned values of $\bqt$ ensuring that both the ingoing and outgoing modes of some field are analytic. Let us first compute a
2-point correlation function. Such a correlator is computed by a pure boundary quantity, and as noted earlier has a one-parameter ambiguity. The fine-tuning involved to make the horizon an ASP is codimension-2 in the $(\omega,k)$ space. Generally, the 2-point correlators are meromorphic with simple poles along the codimension-1 quasinormal spectral curve  $\omega= \omega_{_\text{QN}}(k,m)$ labeled by a discrete parameter $m$.  To determine the behaviour of the boundary Green's function at isolated points associated with our fine-tuning, we need to specify a direction of approach, which gives us our one-parameter ambiguity. Curiously, at these apparent quasinormal modes, examining the 2-point function, one might conclude that the black hole does not Hawking radiate and  appears to lose its thermal character.

Now consider a higher-point function arising from a bulk contact interaction without a horizon pole. Generically, the integrands have a branch-cut, coming from the outgoing Green's function, the coefficient of the advanced sources $\JP$. So the integrand for an $n$-point of the form $\text{F}\ldots \text{FP}\ldots \text{P}$ with $(n-l)$ F and $l$ P fields, can be arranged to not have a branch cut if all the P fields are at a Matsubara ASP point in parameter space. Note that with sufficient retarded (F) fields we do not have a constraint from momentum or frequency conservation. Owing to the absence of a cut in the integrand, such correlation functions vanish. This is not surprising per se; as already noted in~\cite{Jana:2020vyx} and explored further in~\cite{Loganayagam:2022zmq}, even in the generic case correlation functions can vanish at Matsubara frequencies. What is special here is that this happens only when the momenta are also fine-tuned. The one exception is when the vertex has a horizon pole as in~\eqref{eq:cubicdervs}. Then the contour picks up the residue at the pole. 

Thus, unlike the 2-point functions, we have just argued that the effect of the Matsubara ASPs on higher-point functions is more benign. At most the correlation function with some advanced operators tuned to special points in the parameter space vanish. 

We can argue likewise for the energy horizon ASP. Now the energy density operator, say $\MZ$ in a neutral plasma, is analytic for a particular frequency $\omega_\star = 2\pi i T$ and momentum $k_\star$. So correlation functions with a number of advanced operators dual to $\MZ$ will have no branch cut at $(\omega_\star,k_\star)$. Correlators of such operators along with some other advanced operators tuned to the Matsubara ASP, and potentially any number of retarded operators, will produce an integrand with no branch cut emanating from the horizon. Such correlators will vanish, unless there is a horizon pole, which allows one to pick up a localized  residue contribution.  Note that argument does not affect the retarded field. Moreover, it also doesn't guarantee an extended domain of analyticity for the corresponding advanced field, since we only control the behaviour at an isolated point.  The advanced correlator of $\MZ(\omega_\star,k_\star)$ continues to be analytic in the lower-half plane, and potentially free of singularities at this codimension-2 locus. 

The fact that the correlation function of advanced energy density operators at $(\omega_\star,k_\star)$ gets only a localized contribution is reminiscent  of the argument of~\cite{Shenker:2014cwa}. They argued for localized contributions for the out-of-time-ordered correlator at the horizon due to high relative boost (in the shockwave approximation). While we are only computing the Schwinger-Keldysh ordered correlators, the assertion is that exactly at the particular locus where the spatio-temporal dependence of the advanced field takes the form of an exponentially growing mode, we pick up a contribution from the horizon. As noted before, this motivated a proposal for an effective field theory for computing out-of-time-order correlators in maximally chaotic systems presented by~\cite{Blake:2017ris}. They however required that the growing mode was part of the retarded field, but as we have seen, there is nothing special for the retarded field at $(\omega_\star,k_\star)$. The special aspects of this locus pertain to the advanced operator, but all they do is force the correlator to vanish for isolated points.  Our analysis, however, does not indicate whether, there is a direct feature associated with these special points in parameter space for out-of-time-order correlators. 
 
%~~~~~~~~~~~~~~~~~~~~~~~~~~~~~~~~~~~~~~~~~~~~~~~
\subsection{Three-point correlator with horizon pole contribution}\label{sec:example}
%~~~~~~~~~~~~~~~~~~~~~~~~~~~~~~~~~~~~~~~~~~~~~~

To illustrate the general considerations of the preceding discussion, we will examine a three-point correlator arising from derivative interactions in the bulk. Our aim is to explicitly capture the contributions from the horizon pole of the vertex function, as described in~\cref{sec:horpole}. To keep things simple, we will evaluate $I_{_\text{FPF}}$ for the interaction~\eqref{eq:cubicFFC} in the BTZ geometry. This is a simple setting, where 3-point functions for non-derivative interactions can be obtained analytically~\cite{Jana:2020vyx} (four-point functions were computed in~\cite{Loganayagam:2022zmq}). 

Let us start with the simplified expression for the FPF correlator~\eqref{eq:S3FPF}. Expanding out the derivatives $\Dz_\pm$ in~\eqref{eq:S3FPF}, we have  
\begin{equation}
\begin{split}
I_{_\text{FPF}}  
&= 
    e^{\beta\omega_2}\oint r\,dr \ \, e^{-\beta\omega_2\zeta} \left[
    r^2 f \dv{\Gin{\phi}(k_1)}{r}\, \dv{\Gin{\phi}(\bar{k}_2)}{r} 
    -2i\left(\omega_1 \,\Gin{\phi}(k_1)\,\dv{\Gin{\phi}(\bar{k}_2)}{r}-(1\leftrightarrow 2)\right)
    \right. \\ 
&\qquad \qquad  
\left.
    + \frac{\mathbf{k}_1\cdot\mathbf{k}_2}{r^2} \, \Gin{\phi}(k_1)\,\Gin{\phi}(\bar{k}_2) +4\,\omega_1\omega_2\, \frac{\Gin{\phi}(k_1)\Gin{\phi}(\bar{k}_2)}{r^2 f} \right]\Gin{\chi}(k_3)\,.
\end{split}
\end{equation}  
The last term in the above expression has an explicit pole at the horizon, while all other terms are manifestly regular (since the ingoing propagators are analytic). Naively, one might expect this to lead to an IR divergence, but as we have argued, based on a local analysis, this is not the case. To evaluate the integral we employ a simple change of variables, $r = r_+/z$, and rewrite the above as 

\begin{equation}
\begin{split}
I_{_\text{FPF}}^{\text{reg}} 
&= 
    -\left(1-e^{2\pi\bwt_2}\right)\int_{0}^{1}\, \frac{dz}{z} \,
     e^{-2\pi\bwt_2\zeta} \left[
    f \dv{\Gin{\phi}(k_1)}{z}\,\dv{\Gin{\phi}(\bar{k}_2)}{z} 
    +2i\left(\bwt_1 \Gin{\phi}(k_1)\,\dv{\Gin{\phi}(\bar{k}_2)}{dz} 
    - (1\leftrightarrow 2)\right) \right. \\ 
&\qquad \qquad \qquad \qquad  
\left.
    + \bqt_1\cdot\bqt_2\,\Gin{\phi}(k_1)\,\Gin{\phi}(\bar{k}_2)
    \right]\Gin{\chi}(k_3)\,,\\ 
I_{_\text{FPF}}^\text{loc} 
&= 
    4\,\bwt_1\bwt_2\, e^{2\pi\bwt_2}\, \oint \frac{dz}{z f}  \, 
        e^{-2\pi\bwt_2\ctor} \,\Gin{\phi}(k_1)\, \Gin{\phi}(\bar{k}_2) \, 
        \Gin{\chi}(k_3)\,.
\end{split}
\end{equation}
The regular piece, $I_{_\text{FPF}}^{\text{reg}} $, has been reduced to an integral over a single sheet, by picking up the discontinuity across the horizon branch cut. This term is evaluated exactly as in~\cite{Jana:2020vyx}.  

The ingoing bulk-to-boundary propagators have analytic expressions in the BTZ geometry. For a minimally coupled scalar primary of conformal dimension $\Delta$,  one has 
\begin{equation}\label{eq:Ginbtz}
\Gin{\Delta}(z,k) = \frac{\Gamma\left(\bpt_+ +\frac{\Delta}{2}\right)\Gamma\left(\bpt_- +\frac{\Delta}{2}\right)}{\Gamma\left(\Delta-1\right)}z^{\Delta}(1+z)^{-i\bwt}\,
\pFqReg{2}{1}{\bpt_+ + \frac{\Delta}{2},\bpt_- + \frac{\Delta}{2}}{1-i\,\bwt}{ 1-z^2},
\end{equation}
where $\pFqReg{2}{1}{a,b}{c}{z}$ is the regularized hypergeometric function, and 
\begin{equation}
\mathfrak{p}_{\pm} = \frac{i}{2}\left(-\bwt\pm\bqt\right) \,, \qquad z = \frac{r_+}{r} \,.
\end{equation}  
Furthermore, in the present case, we have 
\begin{equation}
e^{-2\pi\bwt\ctor} = \left(\frac{1-z}{1+z}\right)^{i\bwt}\,,
\end{equation}

To evaluate the integrals, we let  $\Delta_1 = \Delta_2 = \Delta_\phi$ and $\Delta_3 = \Delta_\chi$, for simplicity, Further introducing,  as in~\cite{Loganayagam:2022zmq}, the function
\begin{equation}\label{eq:Kindef}
\begin{split}
\gfn{}(k,\Delta) 
&\equiv 
   \Gamma\left(\bpt_++\frac{\Delta}{2}\right) \, \Gamma\left(\bpt_-+\frac{\Delta}{2}\right) \, \Gamma\left(1 - \Delta\right),  
\end{split}
\end{equation} 
in terms of which, the boundary retarded Green's function of a primary with dimension $\Delta$ is given by
\begin{equation}
\Kin{\Delta}(k) 
=
    2 \left(\Delta - 1\right)\, 
    \frac{\gfn{}(k, \Delta)}{\gfn{}(k,2-\Delta)} \,.
\end{equation}  
The rest of the evaluation proceeds by employing the Mellin-Barnes representation of the hypergeometric function.

For the part of the integrand which is regular on the horizon, we find
\begin{equation}
\begin{split}
I_{_\text{FPF}}^{\text{reg}} 
&= 
    -\frac{1}{2}\left(1-e^{2\pi\bwt_2}\right)
    \Gamma\left(1+i\bwt_2\right)\, \Kin{\Delta_{\phi}}(k_1)\, \Kin{\Delta_{\phi}}(\bar{k}_2)\, \Kin{\Delta_{\chi}}(k_3)\, \mathfrak{J}_{_\text{reg}}\,.
\end{split}
\end{equation}
We have written the expression by factoring out the boundary two-point function for the fields, which makes manifest, as in~\cite{Jana:2020vyx,Loganayagam:2022zmq} the analytic structure of the correlators, for the form factor $ \mathfrak{J}_{_\text{reg}}$ is an analytic function in the Fourier domain. Since the  expression for this function is a bit cumbersome, it is relegated to~\cref{sec:Jfns}.

The localized contribution, which is new, instead gives upon using the Mellin-Barnes representation, the following:
\begin{equation}
\begin{split}
I_{_\text{FPF}}^{\text{loc}}
&=
    -2\,\bwt_1\left(1-e^{2\pi\bwt_2}\right)
    \Kin{\Delta_{\phi}}(k_1) \, \Kin{\Delta_{\phi}}(\bar{k}_2)
     \, \Kin{\Delta_{\chi}}(k_3)
    \\ 
&\quad  
   \left(\prod_{i=1}^{3}
    \int_{\mathcal{C}_i}\frac{ds_i}{2\pi i}\, \Gamma(s_i)\, \frac{\Gamma(1-\Delta_i+s_i)}{\Gamma\left(1-\Delta_i+2s_i\right)}\right) \frac{\gfn{}(k_1,\Delta_{\phi}-2s_1)}{\gfn{}(k_1,\Delta_{\phi})}\frac{\gfn{}(\bar{k}_2,\Delta_{\phi}-2s_2)}{\gfn{}(\bar{k}_2,\Delta_{\phi})}
    \frac{\gfn{}(k_3,\Delta_{\chi}-2s_3)}{\gfn{}(k_3,\Delta_{\chi})}\\ 
&\qquad 
    \times
    \left[2\,\bwt_2 \int_{0}^{1-\epsilon} \,dz \,  (1-z^2)^{i\bwt_2-1}\, z^{\sum_{i=1}^{3}(\Delta_i-2s_i)-1} -2^{i\bwt_2}\epsilon^{i\bwt_2}
\right]
\,.
\end{split}
\end{equation}
The key point to note is that the IR divergence from the radial integral, precisely cancels against the localized contribution (the last term in the final line). This is indeed in accord from the local analysis of~\cref{sec:horpole}. Evaluating the integral, we find at the end of the day, 
\begin{equation}
I_{_\text{FPF}}^{\text{loc}}
= 
    2 i\,\bwt_1\left(1-e^{2\pi\bwt_2}\right) \, \Gamma(1+i\bwt_2)\, 
    \Kin{\Delta_{\phi}}(k_1) \, \Kin{\Delta_{\phi}}(\bar{k}_2)
     \, \Kin{\Delta_{\chi}}(k_3) \, \mathfrak{J}_{_\text{loc}}
\end{equation}  
The form factor $ \mathfrak{J}_{_\text{loc}}$, which is also given in~\cref{sec:Jfns}, is manifestly regular. Thus, all the singularities of the three-point function are isolated in the pre-factors, which are products of two-point functions. Combing the two pieces, we find
\begin{equation}
I_{_\text{FPF}} = 
-\frac{1}{2}\left(1-e^{2\pi\bwt_2}\right)
    \Gamma\left(1+i\bwt_2\right)\, \Kin{\Delta_{\phi}}(k_1)\, \Kin{\Delta_{\phi}}(\bar{k}_2)\, \Kin{\Delta_{\chi}}(k_3)\, \mathfrak{J} \,.
\end{equation}  
The form factor $\mathfrak{J}$ can be found in~\eqref{eq:Jformfinal}.

This result is similar to that obtained in~\cite{Jana:2020vyx,Loganayagam:2022zmq}, where there were no vertices with poles at the horizon. In particular, the analytic structure of the three-point function is unchanged from the term containing the horizon pole. There are quasinormal modes for the two F operators in the lower half of the $\omega_1$ and $\omega_3$ plane, and anti-quasinormal modes in the upper half $\omega_3$ plane. This piece has been isolated in the first line (which can be rewritten in terms of the product of the two-point function of the three fields). There are some zeros of the correlator at Matsubara frequencies, from the Boltzmann factor $\left(1-e^{2\pi\bwt_2}\right)$ (though half of these cancel the poles of $\Gamma(1+i\,\bwt_2)$). The reader can also verify that there are no features due to the apparent singularities at the horizon, as argued in~\cref{sec:apphor}. 

While this example illustrates how the grSK contour deals with singularities at the horizon, writing down an example with the $\Lk$ terms in the vertex is a bit more involved. We have however identified a situation, involving the Chern-Simons coupling of a $U(1)$ gauge field in Einstein-Maxwell-Chern-Simons theory, where the computation of anomaly induced terms in the thermal correlator requires understanding both the horizon poles, discussed herein, and the potential singularities at the zeros of $\Lk$. This analysis will appear in a separate publication. 

%~~~~~~~~~~~~~~~~~~~~~~~~~~~~~~~~~~~~~~~~~~~~~~~
\section{Discussion}
\label{sec:discuss}
%~~~~~~~~~~~~~~~~~~~~~~~~~~~~~~~~~~~~~~~~~~~~~~ 

Our goal in this paper was to provide further evidence that real-time thermal correlation functions in holographic field theories are sensibly computed on the grSK geometry introduced in~\cite{Glorioso:2018mmw}. While it has been clear from earlier works~\cite{Chakrabarty:2019aeu,Jana:2020vyx,Loganayagam:2022zmq} that the prescription determines the correlators consistent with Schwinger-Keldysh and KMS conditions, there were some specific features associated with conserved currents that had not been addressed hitherto. Most of the features we discussed are not always directly visible in a low-energy gradient expansion, which owing to the absence of closed form solutions, we generically have to resort to for analytic expressions.

A related motivation for us was to close the gaps in our recent discussion regarding the analytic structure of holographic thermal correlators~\cite{Loganayagam:2022zmq}. There we had assumed that the bulk vertices did not have any singularities which interfered with the grSK contour. As we have seen here, there are indeed such vertices, either when the bulk fields have derivative interactions, or when we consider the gauge invariant combination that corresponds to the energy density field (and charge diffusion field at finite density). 

One could have argued that such singularities, especially the latter, might have originated from our poor choice of variables. However, for reasons described in~\cite{Ghosh:2020lel} (and amplified in~\cite{He:2022jnc,He:2021jna,He:2022deg} for charged plasmas) the bulk fields we have worked with are natural. They correspond to gauge invariant data, have good time-reversal transformation properties, all of which allows for determination of bulk Green's functions without ambiguities. The price one pays is the relative complexity of the differential equation obeyed by the fields. Therefore, the first task we undertook here was to clarify the nature of the differential equations involved and elucidating their general structure. 

We find it curious that all the differential equations of import in static, translationally invariant AdS black hole backgrounds analyzed here are Fuchsian. This has been known for the massive scalar wave equations~\cite{Aminov:2020yma} (and in fact known already in ~\cite{Horowitz:1999jd} for massless scalars), and we verify it to be so even for conserved currents. While we don't a-priori see a rationale for this structure, as argued in the preceding reference one could relate the connection problem for finding quasinormal modes to  instanton partition functions in supersymmetric field theory. This has indeed been successfully used to extract thermal correlators of scalar primary operators of arbitrary dimension in~\cite{Dodelson:2022yvn} in 4 dimensional holographic CFTs (and generalized to include charge and rotation in~\cite{Bhatta:2022wga}). These analyses work with a different variable $z(r^2)$ that reduces the number of regular singular points to four, allowing for use of known data for Heun equations. Since we wished to use the grSK contour in the radial variable, we have refrained here from adopting such a change of variables in our analysis. It is clear that these techniques can be adapted to deduce current correlators as well. 

One corollary of our analysis was a refined understanding of the purported connection between energy transport and maximal scrambling behaviour of black holes. The effective field theory description of maximally chaotic systems studied in~\cite{Blake:2017ris,Haehl:2018izb} has been justified in holographic context by invoking the phenomena of `pole-skipping'. As has been argued by others, this phenomenon is not just restricted to the energy density operator, but acquires special import in this case, since the wave function in the boundary direction resembles a scrambling mode: a temporally growing mode, which is damped out spatially at a rate fixed by the butterfly velocity. We prefer to view this phenomenon in terms of apparent quasinormal modes (for reasons already explained in~\cite{Loganayagam:2022zmq}). In particular, the mode in question arises when the zeros of a particular function  $\Lk$, that naturally appears in the bulk ODE, coincides with the location of the horizon. 

While for a neutral black hole the apparent quasinormal mode behaves like a scrambling mode, it fails to do so for a charged black hole past a critical value of charge. The specific value of charge does not appear to be singled out from the geometry in any way apart from being well away from extremality. Past this critical charge there is no mode that has exponential growth in time in the energy-momentum tensor components. This suggests to us that the link between such apparent quasinormal modes of the energy density and chaotic dynamics is perhaps coincidental.  

Finally, our analysis also revealed one further surprise for charged black holes. While the black hole singularity is a regular singular point for all the fields analyzed herein in a neutral black hole background, it generically isn't for the charged black hole. Per se, this is not surprising, since the causal structure of the two spacetimes are different. However, it is curious that the momentum diffusion  and the transverse photon fields around a \RNAdS{}
black hole actually are sensitive to the black hole singularity. This deserves further investigation, to understand whether such operators can be used to extract information about the spacetime in the vicinity of the singularity along the lines investigated in~\cite{Fidkowski:2003nf,Festuccia:2008zx,Grinberg:2020fdj}.

%~~~~~~~~~~~~~~~~~~~~~~~~~~~~~~~~
\section*{Acknowledgements}
%~~~~~~~~~~~~~~~~~~~~~~~~~~~~~~~~

We would like to thank Lorenz Eberhardt, Felix Haehl, Tom Hartman, Veronika Hubeny, Godwin Martin,  Mark Mezei, and Omkar Shetye for  valuable discussions. We also would like to thank an anonymous referee for suggestions and improvements to the presentation. 
RL acknowledges support of the Department of Atomic Energy, Government of India, under project no. RTI4001, and would also like to acknowledge his debt to the people of India for their steady and generous support to research in the basic sciences.  MR would also like to thank ICTS, Bengaluru, and TIFR, Mumbai for hospitality during the course of this project.  MR was supported  by U.S.\ Department of Energy grant DE-SC0009999 and funds from the University of California. JV was supported by  U.S. Department of Energy grant DE-SC0020360 under the HEP-QIS QuantISED program.
\appendix

%~~~~~~~~~~~~~~~~~~~~~~~~~~~~~~~~~~~~~~~~~~~~~~~
\section{A compilation of wave equations}
\label{sec:eqcompilation}
%~~~~~~~~~~~~~~~~~~~~~~~~~~~~~~~~~~~~~~~~~~~~~~

We compile here the set of equations that have been analyzed hitherto in the context of scalar, Maxwell, and graviton perturbations in \AdS{d+1} black hole backgrounds. Broadly speaking the equations of interest reduce to those of a massless scalar equation with non-trivial kinetic terms, and a radially varying potential, schematically of the form given in~\eqref{eq:homeqn}. We will present the equations as they were introduced in the analysis of thermal correlators in~\cite{Ghosh:2020lel,He:2022jnc} for the \SAdS{d+1} perturbations, and in~\cite{He:2021jna,He:2022deg} for the \RNAdS{d+1} case. 
We will write them in a time-reversal invariant form, using the ingoing adapted radial derivative
\begin{equation}\label{eq:Dzdef}
\Dz_+   \equiv  r^2f\, \dv{r} - i\,\omega \,.
\end{equation}  
%

%~~~~~~~~~~~~~~~~~~~~~~~~~~~~~~~~~~~~~~~~~~~~~~~
\subsection{Perturbations of \texorpdfstring{\SAdS{d+1}}{SAdS}} 
\label{sec:sadseqs}
%~~~~~~~~~~~~~~~~~~~~~~~~~~~~~~~~~~~~~~~~~~~~~~ 
The class of equations we need to analyze for the perturbations of planar \SAdS{d+1} with the geometric data given in~\eqref{eq:schwf} are as follows:

\paragraph{Designer scalar equation:} 
Massive designer scalar with index $\ann$, which is captured by the ODE
\begin{equation}\label{eq:Mdesign}
\begin{split}
\frac{1}{r^\ann} \Dz_+  \left( r^\ann\, \mathbb{D}_+\sen{\ann}\right)+\left(\omega^2-k^2f -  m^2\,r^2 f \right)\, \sen{} =0\,.
\end{split}
\end{equation}
The interesting cases here are $\ann =d-1$ and $m^2 \neq 0$ for a regular massive scalar, dual to a conformal primary of weight $\Delta$ of the boundary CFT. A special case is the massless scalar, which is also the equation for transverse tensor gravitons (which comprises of $\frac{1}{2}\, d\, (d-3)$ polarizations).  In addition, $\ann=1-d$ gives the non-Markovian modes of momentum diffusion (with $d-2$ polarizations), arising from vector perturbations of the metric.  Probe Maxwell fields have modes with $\ann =d-3$ for vector polarizations and $\ann = 3-d$ for scalar polarizations~\cite{Ghosh:2020lel}.

\paragraph{Scalar graviton perturbations:} 
The one other equation of interest is the scalar graviton polarization, which characterizes the dynamics of the energy density field, and takes the form~\cite{He:2022jnc}
\begin{equation}\label{eq:Zsound}
\begin{split}
r^{d-3}\,\Lk(r)^2 \, \Dz_+
        \left( \frac{1}{r^{d-3}\,\Lk(r)^2} \, \Dz_+  \MZ \right)    
            + \left(\omega^2 - k^2 f  \left[1-\frac{d\,(d-2)\, r_+^d}{r^{d-2}\, \Lk(r)}\right]\right) \MZ  =0 \,,
\end{split}
\end{equation}  
with $\Lk$ given in~\eqref{eq:Lkdef}, reproduced here for convenience
\begin{equation}\label{eq:LkdefA}
\Lk(r) = k^2 + \frac{d-1}{2}\, r^3\, f'(r)\,.
\end{equation}  

These equations have been analyzed in \cref{sec:designerscalar} and \cref{sec:energyLk}, respectively.

%~~~~~~~~~~~~~~~~~~~~~~~~~~~~~~~~~~~~~~~~~~~~~~~
\subsection{Perturbations of \texorpdfstring{\RNAdS{d+1}}{RNAdS}} 
\label{sec:rnadseqs}
%~~~~~~~~~~~~~~~~~~~~~~~~~~~~~~~~~~~~~~~~~~~~~~ 

The class of equations we analyze for the perturbations of planar \RNAdS{d+1} with line element~\eqref{eq:fRN} are the following:

\paragraph{Designer scalars:}  The perturbations in the tensor sector, or generic massive perturbations of the black hole are captured by~\eqref{eq:Mdesign} with the replacement of $f(r)$ to~\eqref{eq:fRN}~\cite{He:2021jna}.

\paragraph{Vector polarizations of gravitons and gauge field:} The vector perturbations of metric and gauge field (parity preserving) are captured by equations for two fields, one dual to the momentum current, and the other dual to transverse photons, denoted $\MX$ and $\MY$, respectively.\footnote{ 
    Each of these fields carries a polarization label $\ai=1,2,\ldots, d-2$ which we do not indicate here.}
These fields have a non-trivial dilatonic modulation of their kinetic terms, and a potential which depends on the spatial momenta. The equations as re-derived in ~\cite{He:2021jna} are 
\begin{equation}\label{eq:RNvector}
\begin{split}
\frac{1}{r^{d-3} \, (1-h)^2} \Dz_+ \left((1-h)^2\, r^{d-3}\, \Dz_+ \MX\right) 
+ r_+^2 \left(\bwt^2 - \bqt^2 f\,+  \bRQ^2\, (1-h) \, \BQTv^2 \, f\right) \MX 
&=0 \,,\\
\frac{1}{r^{d-3} \,h^2} \Dz_+ \left(h^2\, r^{d-3}\, \Dz_+ \MY \right) 
+ r_+^2 \left(\bwt^2 - \bqt^2 f\, - \bRQ^2\, (1-h) \, \BQTv^2\, f\right) \MY 
&=0 \,.
\end{split}
\end{equation}
The parameter $\bRQ$ is defined as 
\begin{equation}\label{eq:CCdef}
\bRQ \equiv \frac{(d-2)\, \mu}{r_+\, \sdc}\,,
\end{equation}  
which in turns defines a deformed spatial momentum variable $\BQTv$
\begin{equation}\label{eq:BQTvdef}
\begin{split}
\BQTv^2 
&=
     \sqrt{1+2\frac{\bqt^2}{\bRQ^2}} - 1  = 
    \frac{\bqt^2}{\bRQ^2} \left(1- \frac{1}{2}\, \frac{\bqt^2}{\bRQ^2} +\cdots \right) \,.
\end{split}
\end{equation}  
Note that we have added a $v$ subscript to this momentum variable (relative to~\cite{He:2021jna}) to distinguish it from the  deformation that appears in the scalar sector. 

\paragraph{Scalar polarizations of gravitons and gauge field:}  The scalar sector equations are a bit more complicated. They are given in terms of two fields $\Vd$ and $\Zd$ which correspond to the charge diffusion and energy density mode, respectively. These two fields obey  the following decoupled equations as re-derived in~\cite{He:2022deg}
\begin{equation}\label{eq:RNscalar}
\begin{split}
&
    r^{d-3}\,h^2\, \Dz_+\left(\frac{1}{r^{d-3}\,h^2}\, \Dz_+\Vd\right) + \left( \omega^2 -  k^2 f   +\VVd\right)  \Vd = 0 \,,\\ 
& 
    \frac{r^{d-3}\,\Lk^2}{h^2}\, \Dz_+\left(\frac{h^2}{r^{d-3}\,\Lk^2}\, \Dz_+\Zd\right) + \left( \omega^2 - \left(1-\frac{(d-2)}{2}\,\frac{2+\BQTs^2}{1+\BQTs^2}\,\frac{r^3 f'}{h\,\Lk}\right) k^2f + \VZd \right) \Zd =0 \,.
\end{split}
\end{equation}
In this case we define the deformed momentum parameter $\BQTs$ (note the subscript) as 
\begin{equation}\label{eq:BQTsdef}
\begin{split}
\BQTs^2 
=  \sqrt{1+2\,d\,\nu_s\, \frac{\bqt^2}{\bRQ^2}}-1\,, \qquad 
\nu_s 
= \frac{2(d-2)}{d(d-1)}\,. 
\end{split}
\end{equation}
Note that the function $\Lk$ continues to be given by the expression~\eqref{eq:Lkdef}. 

The potentials are complicated expressions involving various combinations of background functions and the deformed momentum.  For the field $\Vd$ we find the following expression for the potential $\VVd$:
\begin{equation}\label{eq:VdPot}
\begin{split}
\VVd 
&= 
    (d-2) \, \frac{k^2\, f}{(1+\BQTs^2)\,\Lk^2}\,(1-h)\; \VVd^{(1)} 
    +(d-2)\, \frac{r^3 f' f}{4\,(1+\BQTs^2)\,h^2\,\Lk^2} \, \BQTs^2\; \VVd^{(2)}\,,\\
\VVd^{(1)} 
&=
    -\frac{4}{d-1}\, \Lk^2+(d-2)\,(d-1)\,r^5 f' f \left(\frac{1-2\,h}{h}\right)^2 - \BQTs^2 \left(2f(h-(d-2))+r f'\,h\right)\frac{r^2\,\Lk}{h^2}\\
&\quad
    +2\left[\left(d-2-\left(1+2(d-3)\,h\right)h\right)f+(2h-1)\,r f'\, h\right]\frac{r^2\,\Lk}{h^2}\, ,\\
\VVd^{(2)} 
&= 
    2\left(2h-1\right)h \,\Lk^2  + 2(d-1)\left[2\,(d-2)\,f-(5d-9)\,f h+\left(4\,(d-2)\,f-r f'\right)h^2\right]
    r^2 \, \Lk\\
&\quad
-(d-2)\,(d-1)^2\,(2h-1)^2\, r^5 f' f\,.
\end{split}
\end{equation}
On the other hand the potential for $\Zd$ takes the form
\begin{equation}\label{eq:ZdPot}
\begin{split}
\VZd 
&= 
    \frac{d-2}{(1+\BQTs^2)\,\Lk^2}\, k^2\, (1-h)f \;  \VZd^{(1)} - \frac{(d-1)\, (d-2)\, r^5 f'}{2\, (1+\BQTs^2)\, \Lk^2\, h^2}\, \BQTs^2f \; \VZd^{(2)}  \,,\\
\VZd^{(1)} 
&= 
    \frac{4}{d-1}\, \Lk^2 + 2\left[2\left((d-3)\, f-r f'\right)h-(d-3)\,(1+\BQTs^2) f\right]
    \frac{r^2\,\Lk}{h} \\
&\quad
    + \frac{(d-1)\,(d-2)\,r^5 f f'}{h^2}\left(1+\BQTs^2-2\,h\right)\left(2\,h-1\right) \,,\\
\VZd^{(2)} 
&= 
    \left((d-3)\, f-\frac{1}{2}r f'\right)h\,\Lk - \frac{1}{2}\, (d-1)\,(d-2)\,r^3 f f'\,(2h-1)\, ,
\end{split}
\end{equation}
As noted earlier, these expressions were derived in~\cite{Kodama:2003kk}, but our presentation and rewriting of them should make the structure more transparent. It is important that the modulation function $\Lk$ only appears in the potential terms in the dynamics of  $\Vd$.

We can rewrite these equations in the standard form~\eqref{eq:SOLDE}. The  coefficient functions for the SOLDE for $\Vd$ are given to be 
\begin{equation}\label{eq:pqVd}
\begin{split}
p_\Vd(r)
&=
    \frac{5-d}{r} + \frac{r^2f' - 2i\,\omega}{r^2\,f} - 2\, \frac{h'}{h} \\
q_\Vd(r)
&= 
    \frac{(d-1)\,(d-2)^2\,r\,(1-2\,h)\,f'}{4\, (1+\BQTs^2)\, h^2\, \Lk^2}\left[
    4\,k^2\,(h-1) +(d-1)\, \BQTs^2\, r^3\,f'
    \right] \\
&\quad
    - \frac{1}{\Lk\, f}\;\frac{(d-2)\,f'}{2\,(1+\BQTs^2)\, r\, h} \left[2\,k^2\, (2+\BQTs^2 -4\,h)\,(1-h) + (d-1)\,\BQTs^2\, r^3\,h\,f'\right] \\
&\quad
    +\frac{d-2}{2\,(1+\BQTs^2)\,r^2\, h^2\Lk}   
    \left[4\,k^2\, (1-h)\left((\BQTs^2+1) \,(d-2-h) - 2\, (d-3)\, h^2\right) \right. \\
&\left. \qquad\qquad 
    + (d-1)\,\BQTs^2\, r^3\,h\,f' \left(2\,(d-2) (2\,h^2+1) + h \, (9-5\,d)\right)
    \right] \\
&\quad
    +\frac{1}{r^3 f\,h} \bigg[
    \frac{2\,k^2\,h \left(9-5\,d - (d-1)\, \BQTs^2 +4\,(d-2)\,h\right) +(d-1)\,(d-2)\, (2\,h-1)\,\BQTs^2\, r^3\,f'}{2\,(d-1)\,(\BQTs^2+1)\,r} \\
& \qquad \qquad \qquad 
    + i\omega\, \left((d-3)\,h+2\,r\,h'\right)    
    \bigg].   
\end{split}
\end{equation}
For the field $\Zd$ we instead find the coefficient functions
\begin{equation}\label{eq:pqZd}
\begin{split}
p_\Zd(r)
&=
    \frac{5-d}{r} + \frac{r^2f' - 2i\,\omega}{r^2\,f} + 2\, \frac{h'}{h} - 2\, \frac{\Lk'}{\Lk} \,, \\
q_\Zd(r)
&= 
    \frac{(d-1)\, (d-2)^2\, r\, (2h-1)\,f'}{4\,\BQTs^2\,h^2\, \Lk^2} \left[
    4\,k^2 \left( 1+\BQTs^2 - (3+\BQTs^2) \,h + 2\,h^2\right) + (d-1)\, \BQTs^2\, r^3\, f'    \right] \\
&\quad
    + \frac{1}{\Lk\,f}\left[\frac{2\Lk'}{r^2} i\omega + \frac{(d-2)\,f'}{4\,(1+\BQTs^2)\, h} \left[2\,k^2 \left(2+\BQTs^2 + 8\,h (1-h)\right) + (d-1)\, \BQTs^2\, r^3\, f'\right]\right] \\
&\quad
    - \frac{(d-2)\,(d-3)}{2\,(1+\BQTs^2)\, r^2\, h\, \Lk}\left[
    4\,k^2 \left(1+\BQTs^2 - (3+\BQTs^2) \, h + 2\, h^2 \right)+(d-1)\, \BQTs^2\, r^3\, f'     \right] \\
& \quad
+   \frac{1}{r^4\,f} \left[
    -2i\,\omega \frac{r^2\,h'}{h}
    + \frac{(d-1)\,(d-3)\,(1+\BQTs^2)\,r \, i\omega -k^2\,(7-3\,d + (d-1) \,\BQTs^2 +4\,(d-2)\,h)}{(d-1)\, (1+\BQTs^2)}
    \right].  
\end{split}
\end{equation}
%

%~~~~~~~~~~~~~~~~~~~~~~~~~~~~~~~~~~~~~~~~~~~~~~~
\section{Scalar metric and gauge field perturbations of the charged black hole}
\label{sec:sRNpertpars}
%~~~~~~~~~~~~~~~~~~~~~~~~~~~~~~~~~~~~~~~~~~~~~~

We had promised in the main text to relate these fields to the physical metric and gauge field perturbations. Writing the metric and gauge field for the Einstein-Maxwell theory as 
\begin{equation}\label{eq:EMpert}
\begin{split}
ds^2 = ds_{(0)}^2 + ds_{(1)}^2 \,, 
\qquad \vb{A} = - a\, dv + \AGR_A\, dx^A \,, 
\end{split}
\end{equation}
with the subscript `$(0)$' referring to the background solution. The metric perturbations are parameterized by fields $\{\PHE,\PHO,\PHW\}$ as in~\eqref{eq:scalarpert}, while the gauge field perturbation is captured by another  field $\MV$ 
\begin{equation}\label{eq:EOWDeb}
\begin{split}
 ds_{(1)}^2
 &=  
    \frac{\PHE-r f\,\PHW}{r^{d-3}}\, dv^2 
     +\frac{2}{r^{d-1}f} \left(\PHO-\PHE +rf\,\PHW\right) dv\,dr
     + r^2\, \frac{\PHW}{r^{d-2}}\, d\vb{x}^2 \\
& \qquad \qquad 
    -\frac{1}{r^{d+1}f^2}\left[2(\PHO-\PHE) + r f \,(d-1)\, \PHW\right] dr^2  \,,\\
\AGR_A\, dx^A 
&=
     \frac{1}{r^{d-3}} \left(dv\, \Dz_+ -  dr \,\dv{r}\right) \MV   \,.
\end{split}
\end{equation}  
These four functions are given in terms of $\Vd$ and $\Zd$ through the following relations (we have chosen not to solve for the fields directly).
First we redefine the metric functions using a field similar to $\MZ$ encountered in the \SAdS{d+1} example, taking into account the mixing between the gauge field and graviton perturbations:
\begin{equation}\label{eq:EOWMZ}
\begin{split}
\PHO
&= 
    -i\omega\, \bwt\,, \qquad 
\PHE 
=   
    \Dz_+ \bwt -2\, a'\,r^2f\, \MV \,,\\ \medskip
\bwt 
&= 
    \frac{r}{\Lk}\left[\Dz_+  -\frac{r^2\, f'}{2}\right]   \MZ + \frac{2\,(d-1)\, r^3f\, a'}{\Lk}\, \MV \,, \\
\PHW 
&= 
       \frac{1}{ \Lk}\left[r\,\Dz_+ + \frac{k^2}{d-1} \right]\, \MZ+ \frac{2\,(d-1)\, r^3f\, a'}{\Lk}\, 
       \MV\,.
\end{split}
\end{equation}  
The fields $\MV$ and $\MZ$ themselves are finally given in terms of $\Vd$ and $\Zd$ by a functional basis change in a set of coupled ODEs
\begin{equation}\label{eq:MZVdiagonal}
\begin{split}
\MZ &= \frac{\Lk}{\bRQ\, h} \, \Vd + h\, \Zd \,, \\
\MV 
&=  
    \left(\frac{d-2}{\bRQ}\, a - r_+\, \frac{\BQTs^2+2}{2}\right) \frac{\Vd}{2\, h} +\left( (d-2) a + \frac{r_+}{2} \, \bRQ\, \BQTs^2\right) \frac{h}{2\, \Lk} \, \Zd \,.
\end{split}
\end{equation}

From our analysis in \cref{sec:sRNads} we noticed that while $\Zd$ is regular at either set of roots $r_{k1}$ and $r_{k2}$ of $\Lk$, the function $\Vd$ was only regular at the former, but had a simple pole at the latter. From here we conclude that $\MZ$ is regular, but $\MV$ potentially has a simple pole. 

However, upon using the local Frobenius solution for $\Zd$ and $\Vd$, we find however that both $\MZ$ and $\MV$ are regular at all the roots of $\Lk$. Furthermore, the fields $\{\PHE,\PHO,\PHW\}$ themselves are regular with the residue at the poles coming from the $\Lk$ factors canceling out. Apart from noting that this is indeed physically sensible -- metric functions and gauge fields should not have divergences -- we do not have an explanation for the conspiracy in the coefficients which achieves this. The best we can offer is to note that we could have anticipated on physical grounds that the function $\Vd$ has to be singular like $\Lk^{-1}$. The fact that it is only singular at the $r_{k2}$ set of roots has to do with the relative coefficients in the linear combination (recall that the roots are controlled by $\BQTs^2$ and $\BQTs^2+2$). 

%~~~~~~~~~~~~~~~~~~~~~~~~~~~~~~~~~~~~~~~~~~~~~~~
\section{Analysis of black hole wave equations}\label{sec:bhwave}
%~~~~~~~~~~~~~~~~~~~~~~~~~~~~~~~~~~~~~~~~~~~~~~

As we undertake our analysis we will have the need to refer to various singular points (abbreviated SP). Some of them will occur for generic values of parameters corresponding to the background geometry and the linearized field, while some appear upon some fine-tuning. We will distinguish the two cases for convenience, isolating also the features that are specific to certain probes. 

%~~~~~~~~~~~~~~~~~~~~~~~~~~~~~~~~~~~~~~~~~~~~~~~
\subsection{Terminology for singular points}
\label{sec:terminology}
%~~~~~~~~~~~~~~~~~~~~~~~~~~~~~~~~~~~~~~~~~~~~~~

The  SPs of interest that appear at generic values of the parameters occur for the most part at geometric loci. The obvious locations are the AdS boundary, the black hole singularity, and zeros of the function $f$ (including the horizon). We find it convenient to refer to these using the following terminology:
\begin{itemize}[wide,left=0pt]
\item  Asymptotic SP, refers to the point at the AdS boundary $r\to \infty$.
\item Curvature SP, associated with the black hole singularity. 
\item Horizon SP, located at the horizon for a non-degenerate black hole, and other roots of the metric function $f$.
\end{itemize}

In addition, there are some singular points that are specific to equations governing conserved currents, after repackaging the information into gauge invariant variables (as described above). We encounter 
\begin{itemize}[wide,left=0pt]
\item Energy density ASP: This is the potential SP from the zero of the function $\Lk$,~\eqref{eq:Lkdef} encountered in the analysis of scalar polarizations of gravitons (gauge fields). 
\item Ohmic SPs: These are specific to the \RNAdS{d+1} solution, and are determined by the Ohmic radius  $\RQ$ introduced in~\cite{He:2021jna} (related to zeros of $f'$) typically lying inside  the black hole horizon. 
\end{itemize}

The SPs listed above exist for generic parameter choices, viz., for generic values of black hole parameters and at generic frequencies and momenta of probes. However, additional features arise as we vary parameters, either in the form of confluence, where two regular SPs merge and become irregular, or coalescence, where the merger remains a regular SP.\footnote{
    The adjective confluence is used traditionally in the theory of differential equations to characterize mergers of singular points which lead to essential singularities in the solution (hence irregular SP). When the merger does not change the nature of the SP we choose to refer to it as a coalescence (even when it changes the characteristic exponents).  
} 
Moreover, either for the geometric SPs or at confluence/coalescence, we may be able to convert the SP to be an ASP by a further fine-tuning of the parameters. For these, our terminology will be the following, defined in decreasing order of genericity:

\begin{itemize}[wide,left=0pt]
\item Extremal confluence: The extremal limit where the horizon SPs become irregular.\footnote{ 
    This may also involve a merger with the Ohmic singular point.}
\item Matsubara ASP: These refer to special fine-tuned values of frequencies, specifically  $\omega = -2\pi i \, n\, T $ with $n \in \mathbb{Z}_{+}$, where the horizon SP has some special features. At this locus the ingoing and outgoing solution both become analytic (at certain specific values of the momenta), resulting in the horizon becoming an ASP. This was noticed in~\cite{Blake:2018leo,Grozdanov:2019uhi} and explored in some generality in~\cite{Blake:2019otz}. 
\item Energy asymptotic coalescence: The equation of the scalar graviton polarization has additional features for translational homogeneous modes 
($k=0$) owing to the presence of soft gravitons~\cite{He:2022jnc}. Here the asymptotic characteristic exponents change, whilst however retaining the regular singular nature of the SOLDE. 
\item Energy horizon coalescence: Refers to the situation when the horizon SP merges with the energy density ASP. This occurs at a codimension-2 locus, at fixed $\omega, k$, and has been analyzed hitherto in the context of chaotic dynamics of black holes~\cite{Blake:2018leo}. 
\end{itemize}

%~~~~~~~~~~~~~~~~~~~~~~~~~~~~~~~~~~~~~~~~~~~~~~~
\subsection{Schwarzschild-AdS black hole wave equations}
\label{sec:schw}
%~~~~~~~~~~~~~~~~~~~~~~~~~~~~~~~~~~~~~~~~~~~~~~

As remarked above, some of the general features we explore below have been uncovered in the literature. There are a few situations primarily involving the charged \RNAdS{d+1} black hole where there is a qualitatively new behaviour. For completeness, however, we will describe the neutral case first, phrasing the results in a manner which resonates with our understanding of the physics.  The background solution is the planar \SAdS{d+1} black hole whose metric takes the form~\eqref{eq:dssq} with 
\begin{equation}\label{eq:schwf}
f(r) = 1-\frac{r_+^d}{r^d}\,, \qquad T = \frac{d\, r_+}{4\pi}\,.
\end{equation}  
%

%~~~~~~~~~~~~~~~~~~~~~~~~~~~~~~~~~~~~~~~~~~~~~~~
\subsubsection{Designer scalar ODEs}
\label{sec:designerscalar}
%~~~~~~~~~~~~~~~~~~~~~~~~~~~~~~~~~~~~~~~~~~~~~~

We begin with the simplest equation~\eqref{eq:SLMdesign} capturing the dynamics of a designer scalar field. It has the geometric singular points at $r=\infty$, $r=0$, and at $f(r) =0$. This comprises $d+2$ singular points for a probe of \SAdS{d+1}, rendering it as a Fuchsian equation with the given number of regular singular points~\cite{Aminov:2020yma}.  The local behaviour of the differential equation in the neighbourhood of the singular points can be summarized as follows:
\begin{itemize}[wide,left=0pt]
\item  Asymptotic SP\footnote{
    We have directly expanded the coefficient functions to determine the behaviour at the asymptotic SP. A more natural way to deduce this  would have employed a fractional linear map $\ri = \frac{r_+}{r} $ to bring the point to the origin. 
} 
\begin{equation}\label{eq:DMasymSP}
p_\varphi = \frac{\ann+2}{r} + \order{r^{-2}} \,,
\qquad
q_\varphi = -\frac{m^2}{r^2} + \order{r^{-1}} \,.
\end{equation}  
\item  Curvature SP
\begin{equation}\label{eq:DMcurvSP}
p_\varphi = \frac{\ann+2-d}{r} + \order{r^{0}} \,,
\qquad
q_\varphi = \bqt^2 \, \frac{r^{d-4}}{r_+^{d-2}} + \order{r^{d-3}} \,.
\end{equation}  
\item Horizon SP\footnote{
    When we write the local behaviour at the horizon SP, we are going to do so only for the largest positive real root of $f$. If we wish to examine the behavior at a different zero of $f$, say $r = r_+ \varpi_d$, for some $d^\textrm{th}$ root of unity, then the expression~\eqref{eq:DMhorSP} holds provided we also rotate our dimensionless frequencies and momenta appropriately, viz., replace $r_+ \to \varpi_d\, r_+$ whilst also defining $(\omega,k) = \varpi_d\, r_+\, (\bwt, \bqt)$. \label{fn:horSPs}
}
\begin{equation}\label{eq:DMhorSP}
p_\varphi = \frac{1- \frac{2}{d}\,i\bwt}{r-r_+} + \cdots 
\,,
\qquad
q_\varphi = -\frac{1}{d\,r_+} \frac{\bqt^2 +m^2+ i\, \ann\, \bwt}{r-r_+} + \cdots\,.
\end{equation}  
\end{itemize}
We have written the result in dimensionless frequency and momenta, $\bwt$ and $\bqt$, natural from a bulk perspective.\footnote{
    For the \SAdS{d+1} solution the two dimensionless frequencies are simply related as $\mathbf{w} = \frac{2}{d}\, \bwt$.
}

The singular points are depicted  in the complex radial plane in 
\cref{fig:designerSchw}.

At the asymptotic SP, the characteristic exponents are determined by
\begin{equation}\label{eq:Deltann}
\alpha (\alpha-1) + \alpha (\ann +2 ) - m^2 = 0  \;\; \Longrightarrow \;\; 
\alpha (\alpha + \ann+1) - \Delta (\ann+1-\Delta) =0\,,
\end{equation}  
where, by parameterizing the mass appropriately in terms of a `dimension' $\Delta$, we have identified the behaviour with that of a scalar field in a \AdS{} spacetime of effective dimension $d_\text{eff} = \ann+1$. This system has been analyzed in~\cite{Loganayagam:2022zmq}.

At the curvature SP, the indicial equation implies that the characteristic exponents are $0$ and $d-1-\ann$. Minimally coupled scalar with $\ann =d-1$ always has a branch cut at the singularity (since both exponents are equal). For $d>3$, designer scalars with $d-1-\ann \in \mathbb{Z}_{+}$ naively could find the origin to be an apparent singularity. Examples of such fields are scalar and vector polarizations of a probe Maxwell field in the background, which have $\ann = 3-d$ and $\ann =d-3$, respectively, and vector graviton polarizations which have $\ann = 1-d$. However, one can check that despite the exponents being integral, there is always a logarithmic branch cut, and thus the black hole singularity is a regular SP of the SOLDE. 

Finally, the horizon SP, are at the zeros of $f$, which depends on the geometry in question.  For the planar \SAdS{d+1} black hole one  has $d$-roots at  $r= r_+ \varpi_d$, where $\varpi_d$ are the $d^\text{th}$ roots of unity.\footnote{
    We use $\varpi_n$ to denote the set of $n^\text{th}$ roots of unity, $\varpi_n =\{e^{2\pi i\frac{m}{n}}, m = 0,1,\ldots,n-1\}$. Often we will simply report the behaviour for $m=0$ as noted in \cref{fn:horSPs}. 
}  
At the horizon SP, the characteristic exponents are $0$ and $i\,\mathbf{w}$. The former is the analytic ingoing mode, while the latter is the outgoing mode. While this suffices for the local solution around the horizon SP, if we were to pick some closed form function representation for the solution (e.g., the hypergeometric function usually employed in the BTZ geometry), one should ensure that any branch cuts inherent there should be run in a direction away from the ray $[r_+, \infty)$. From these results, it follows that~\eqref{eq:SLMdesign} is indeed a Fuchsian SOLDE with $d+2$ regular SP. We recall here the Fuchsian sum rule, which demands that the sum of the characteristic exponents at all the SPs equal two less the number of SPs. 

\paragraph{Matsubara ASP:} If we consider the horizon SP, for $i\mathbf{w}=n$ with  $n\in \mathbb{Z}_{\geq 0}$, then the second mode function, viz., the outgoing mode,  can also be made regular at the horizon (this was noticed a long time ago in~\cite{Miranda:2005qx,Miranda:2008vb}).
With this choice we make the two characteristic exponents non-negative integers, but a further fine-tuning is required to ensure the absence of any monodromy. This was explored in~\cite{Blake:2019otz}, and can be understood as follows. The Frobenius expansion for the outgoing mode despite starting out with an integral power, will generically include a logarithmic branch, rendering the mode non-analytic. This can be avoided, since we can tune the momentum $\bqt$. Assuming one of the characteristic exponents to be zero (which can be done by factoring out a power), and the difference $n \geq 1$, we will need to Laurent expand $p_\varphi$ and $q_\varphi$ to order $n-2$. The coefficient $\gamma$ of the logarithmic mode can be determined in terms of these Laurent coefficients, see  \cref{sec:localapparent} for details. Thus, setting a particular combination of the coefficients to vanish will suffice to ensure that we have an ASP.  

For the horizon SP we have, for given $n$, a polynomial of degree $2n$ in 
$\bqt$ determining the non-trivial monodromy. Choosing $\bqt$ to be a root of this polynomial will ensure that we have a horizon ASP.  Once this is done, there are no quasinormal modes, as there is no quantization condition to solve for (we have fixed both $\bwt$ and $\bqt$); imposing normalizability at the boundary is ineffective since any linear combination of the modes on the horizon is acceptable. This is the reason why we have apparent quasinormal modes. Likewise, attempting to find the boundary Green's function leads to an ambiguity. This has been explained in~\cite{Blake:2018leo, Grozdanov:2019uhi,Blake:2019otz,Natsuume:2019xcy}, so we will not elaborate further. The reader may find wave equations on the BTZ background studied recently~\cite{Loganayagam:2022zmq} helpful to see how the ambiguity arises (since analytic expressions are available in this case).

%~~~~~~~~~~~~~~~~~~~~~~~~~~~~~~~~~~~~~~~~~~~~~~~
\subsubsection{The energy density operator ODE}
\label{sec:energyLk}
%~~~~~~~~~~~~~~~~~~~~~~~~~~~~~~~~~~~~~~~~~~~~~~

Let us now look at~\eqref{eq:Zsound} for the planar \SAdS{d+1} black hole. This has been examined in $d=4$ in Appendix B of~\cite{Blake:2018leo} for some particular aspects, but we will highlight some additional features, which will be of relevance to our discussion. The differential equation is for a variable $\MZ(r)$, with coefficient functions 
\begin{equation}\label{eq:ZSLfns}
\begin{split}
p_\MZ(r)
&= 
    \frac{5-d}{r} + \frac{r^2\, f'-2i\omega}{r^2\,f} - 2\,\frac{\Lk'}{\Lk} \,, \\
q_\MZ(r)
&=
    \frac{d\,(d-2)\, r^2\, k^2 + r^d (-k^2 + i \omega \,(d-3) r) \, \Lk +2i\omega\, r^{d+2}\, \Lk'}{r^{d+4}\, f\, \Lk} \,.
\end{split}
\end{equation}  
The local behaviour of the differential equation in the neighbourhood of the geometric singular points can be summarized as follows:
\begin{itemize}[wide,left=0pt]
\item  Asymptotic SP
\begin{equation}\label{eq:ZasymSP}
p_\MZ = \frac{5-d}{r} + \order{r^{-2}} \,,
\qquad
q_\MZ =  \order{r^{-3}} \,.
\end{equation}  
\item  Curvature SP
\begin{equation}\label{eq:ZcurvSP}
p_\MZ = \frac{1}{r} + \order{r^{0}} \,,
\qquad
q_\MZ = \frac{1}{r} + \order{r^{0}} \,.
\end{equation}  
\item Horizon SP
\begin{equation}\label{eq:ZhorSP}
p_\MZ = \frac{1- \frac{2}{d}\,i\bwt}{r-r_+} + \cdots 
\,,
\qquad
q_\MZ = \frac{q_{\MZ,-1}}{r-r_+} + \cdots\,.
\end{equation}  
The residue of $q_\MZ$ at the horizon is 
\begin{equation}\label{eq:qZhorres}
q_{\MZ,-1} = \frac{1}{d\, \Lk(r_+) \, r_+^{d-1}} \left(
    d\, (d-2)\, \bqt^2 - r_+^{d-2} \bqt^2 \Lk(r_+) 
    - \frac{i\bwt}{2} \, r_+^d\, \left[d\,(d-1)^2-2\, (d-3)\, \bqt^2\right] \right).
\end{equation}  
While this particular value is not relevant for the purposes of computing the characteristic exponents, it will be useful when we consider fine-tuned values of parameters.
\end{itemize}

From the asymptotic SP the characteristic exponents for $\MZ$ for generic $(\omega,k)$ are $0$ and $d-4$, respectively, which was used in~\cite{He:2022jnc} to argue for its non-Markovian character. At the curvature SP, we have a degenerate pair of exponents (both zero), which implies that there is a logarithmic branch in the solution for $\MZ$ near the black hole singularity. The behaviour at the horizon is characterized by an ingoing mode that is analytic (exponent $0$) and an outgoing mode which is not -- due to the exponent being $i\mathbf{w}$ -- as is the case for any non-extremal black hole at generic values of frequency and momenta. 

\paragraph{Energy density ASP:} The zeros of $\Lk$ are potential SPs of the equation for $\MZ(r)$. They are located at $r_k$, given in~\cref{eq:rkschw}, and turn out to be ASPs.

To see this we note that the characteristic exponents are $0$ and $3$, determined by the fact that $p_\MZ \sim -\frac{2}{r-r_k}$, while $q_\MZ$ starts off with a simple pole. Working out the coefficients to $\order{r-r_k}$ we find that the relation obtained in~\eqref{eq:nologcond} always holds.  Hence, as advertised in \cref{sec:terminology} it is appropriate to refer to the roots of $\Lk$ as energy density ASP. Thus, the singular points of~\eqref{eq:Zsound} are $d+2$ in number, coinciding for generic $(\omega,k)$ with the picture we had for the designer scalar wave equations \cref{fig:designerSchw}. Taking the energy density ASPs into account, we find the sum of the characteristic exponents to be $4-d + 3\, (d-2) = 2\,d-2$, which is indeed what is demanded by the Fuchsian condition for an equation with $d+2$ regular SPs and $(d-2)$  ASPs.

There are additional features at fine-tuned values of the parameters. We again have a discretuum set of Matsubara ASPs. Here we still tune $\mathbf{w} = -i\,n$ with $n \in \mathbb{Z}_{\geq 0}$. The tuning of $k$ involves also the function $\Lk$ leading to a more complicated condition than before, but one can ascertain that suitable choices of $k$ can be made. Their physical interpretation follows along the lines described in \cref{sec:designerscalar}, so we will not elaborate further. Fine-tuning $k$ to particular values, however, leads to some new features that are unique to the energy density ODE. We will describe two aspects of coalescence that are interesting below.

\paragraph{Energy asymptotic coalescence:} While modes with $k^2 \neq 0$ have the asymptotic SP determined by~\eqref{eq:ZasymSP}, translationally invariant modes, with $k^2 =0$, have a very different behaviour. This has to do with the function $\Lk(r)$ which has non-trivial momentum dependence. The explicit 
$k^2$ dependence in the potential does not affect this argument just as the momentum dependence does not enter the asymptotic fall-offs of fields in \AdS{} spacetimes. In any event, setting $k^2=0$ we find the locus $r=\infty$ is still a regular SP, albeit with a modified set of characteristic exponents, $r^0$ and $r^{-d}$, respectively. This behaviour is akin to that of a minimally coupled, massless scalar field in the geometry.  The Fuchsian nature of the equation at this locus is more obvious, reducing as it does to the earlier analysis of designer scalar equation.

\paragraph{Energy horizon coalescence:} We have a set horizon SPs arising from  the roots of $f$, and a set of energy ASP from $\Lk$. We can ask when two of these coalesce. As with the energy asymptotic coalescence, it is reasonable to expect that the horizon remains a regular SP, perhaps with some changed exponents. Focusing specifically on the largest real zero of $f$, we see that  $\Lk$ will have a root at $r_+$ provided we set
\begin{equation}\label{eq:qLk}
k^2 + \frac{d-1}{2} \, r_+^3\, f'(r_+) = 0  \;\; \Longrightarrow \;\;
\bqt = \pm i \, \sqrt{2\pi \,(d-1)\,r_+\,T}.
\end{equation}  
With this choice we then notice that the coefficient functions become
\begin{equation}\label{eq:pqZks}
\begin{split}
p_\MZ(r) 
&= 
    -\frac{1-i\,\mathbf{w}}{r-r_+} +  \order{(r-r_+)^0} \,, \\ 
q_\MZ(r) &= 
    \frac{1 + i\,\mathbf{w}}{(r-r_+)^2} + \frac{(d-5)\, (1+i\mathbf{w})}{2\, (r-r_+)} + \order{(r-r_+)^0}\,.
\end{split}
\end{equation}  
The main change here is the double pole in $q_\MZ$ which arises from the merger of the two zeros (there is also a shift in the residue of $p_\MZ$). 
The indicial equation for the power law $(r-r_+)^\alpha$,  near the singular point, now indicates that the characteristic exponents shift to 
$\alpha_+ =1$ and $\alpha_- = 1  + i\mathbf{w}$. This is all that happens for generic frequencies. 

However, one might imagine that as with the Matsubara ASP, we can tune the frequency to make the $\alpha_-$ mode regular at the horizon, say by choosing 
$\mathbf{w}= -i\, (n-1)$ for $ n \in \mathbb{Z}_{\geq 0}$. Note however, that the momentum has been fixed already in~\eqref{eq:qLk}, which implies that we don't have too much fine-tuning freedom. Not only, do we have to make sure that the characteristic exponent is integral, but we also have to check that the Laurent coefficients work out to remove the logarithmic branch.  This turns out to be possible, but only at a particular frequency,  $\mathbf{w}= i$, i.e., $\omega = 2\pi i T$. At this point, both modes are analytic, and the horizon locus is an ASP, as we can already see from~\eqref{eq:pqZks}.  For any other choice of $n$, the solutions have a logarithmic branch cut, and the outgoing solution picks up a monodromy.

%~~~~~~~~~~~~~~~~~~~~~~~~~~~~~~~~~~~~~~~~~~~~~~~
\subsection{Reissner-Nordstr\"om-AdS black hole wave equations}
\label{sec:rnadswave}
%~~~~~~~~~~~~~~~~~~~~~~~~~~~~~~~~~~~~~~~~~~~~~~

We now turn to the discussion of wave equations in a charged black hole background corresponding to a boundary CFT plasma at non-zero chemical potential. The background line-element is still given by the translationally invariant~\eqref{eq:dssq}, now with 
\begin{equation}\label{eq:fRN}
f(r) = 1 - (1+Q^2)\,\left(\frac{r_+}{r}\right)^d + Q^2\, \left(\frac{r_+}{r}\right)^{2(d-1)} \,.
\end{equation}  
The physical parameters, temperature and chemical potential, are related to $r_+$ and $Q$ through 
\begin{equation}\label{eq:TmuRN}
\begin{split}
T 
&=  
    \frac{d -(d-2)\, Q^2}{4\pi}\, r_+  \,, \qquad \mu =\sqrt{\frac{d-1}{d-2}} \, Q \, r_+  \,. 
\end{split}
\end{equation}  
We also introduce the Ohmic radius $\RQ$, Ohmic function $h$, and the  dc conductivity parameter $\sdc$  following~\cite{He:2021jna}, for they will play an important role in our analysis. These are given by
\begin{equation}\label{eq:hRQdef}
\RQ^{d-2} = \frac{d-1}{d}\, \frac{2\,Q^2}{1+Q^2}   \, r_+^{d-2}  \,, \qquad 
h(r) =  1-\frac{\RQ^{d-2}}{r^{d-2}} = 1-\sdc\, \frac{r_+^{d-2}}{r^{d-2}} \,,
\end{equation}  
with  
\begin{equation}\label{eq:sdcdef}
\sdc \equiv  \frac{\RQ^{d-2}}{r_+^{d-2}} \,.
\end{equation}  
The wave equations of interest are collected in \cref{sec:rnadseqs}. 

%~~~~~~~~~~~~~~~~~~~~~~~~~~~~~~~~~~~~~~~~~~~~~~~
\subsubsection{Designer scalar ODEs}
\label{sec:desMRN}
%~~~~~~~~~~~~~~~~~~~~~~~~~~~~~~~~~~~~~~~~~~~~~~ 

The designer scalar ODE in the \RNAdS{d+1} black hole background is similar to the that studied in \cref{sec:designerscalar}. It is relevant for the study of massive scalar probes, tensor graviton perturbations of the black hole (which again are massless minimally coupled scalars), both of which have $\ann =d-1$. We will however analyze the equations for arbitrary $\ann$.  

The analysis of the designer ODE~\eqref{eq:SLMdesign} in the  planar \RNAdS{d+1} is similar to our earlier discussion. The key difference is that the function $f(r)$ has $2(d-1)$ zeros from~\eqref{eq:fRN}. Let us first consider a non-degenerate horizon where the roots of $f(r)$, and in particular the largest real-root, are non-degenerate. The asymptotic SP is unaffected by the charge, and the analysis at the horizon SP is similar to the Schwarzschild case (accounting for the charge dependence of the temperature). Specifically, 
\begin{itemize}[wide,left=0pt]
\item  Asymptotic SP 
\begin{equation}\label{eq:DMasymSPRN}
p_\varphi = \frac{\ann+2}{r} + \order{r^{-2}} \,,
\qquad
q_\varphi = -\frac{m^2}{r^2} + \order{r^{-1}} \,.
\end{equation}  
\item  Curvature SP\footnote{
    We are assuming the limit is taken with $Q\neq 0$. The order of limits is important, as higher terms in the Laurent expansions have singular behaviour in the limit $Q\to 0$ (which makes sense since the function $f$ diverges more rapidly for the charged solution). 
}
\begin{equation}\label{eq:DMcurvSPRN}
p_\varphi = \frac{\ann+4-2\,d}{r} + \order{r^{0}} \,,
\qquad
q_\varphi = \order{r^{2(d-3)}} \,.
\end{equation}  
\item Horizon SP
\begin{equation}\label{eq:DMhorSPRN}
p_\varphi = \frac{1- i\mathbf{w}}{r-r_+} + \cdots 
\,,
\qquad
q_\varphi = -\frac{1}{r-r_+} \left(\frac{\beta}{4\pi}\, (\bqt^2 +m^2) + \frac{\ann\, }{2r_+} \, i\mathbf{w} \right)  + \cdots\,.
\end{equation}  
\end{itemize}
In this case using the Matsubara normalized frequency 
\begin{equation}\label{eq:MatRN}
\mathbf{w} = \frac{\beta\omega}{2\pi} \,,
\end{equation}  
results in cleaner expressions, since $T(r_+)$ is more involved.\footnote{
    Now $\mathbf{w}$ and $\bwt$ are related by a factor which is $\frac{d-(d-2)\,Q^2}{2}$, which we prefer to avoid writing, and have hence chosen to work with both normalizations.
} The reader can check that~\eqref{eq:DMhorSP} and~\eqref{eq:DMhorSPRN} are the same accounting for the translation between horizon size and temperature. Note that we are again focusing on the largest positive root of $f$ to define the horizon SP behaviour. As indicated in \cref{fn:horSPs} we can carry out an analogous exercise for the other roots, but we now will have to work with the locally defined (complex) temperatures if we wish to write the expressions in terms of  the dimensionless variables. 

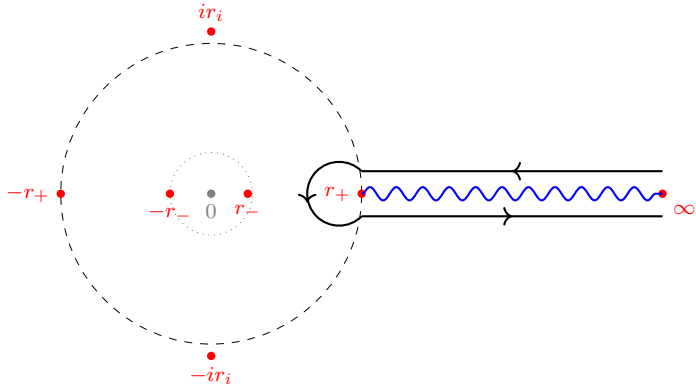
\begin{figure}[ht!]
\centering
\begin{tikzpicture}[scale=1]
\draw[very thin,color=black,dashed] (-5,0) circle (2cm);
\draw[very thin,color=black,dotted] (-5,0) circle (0.54858cm);
\draw[thick,color=gray,fill=gray] (-5,0) circle (0.25ex) node[below] {$\scriptstyle{0}$};
\draw[thick,color=red,fill=red] (-7,0) circle (0.25ex) node[left] {$\scriptstyle{-r_+}$};
\draw[thick,color=red,fill=red] (-3,0) circle (0.25ex) node[left] {$\scriptstyle{r_+}$};
\draw[thick,color=red,fill=red] (-5.54858,0) circle (0.25ex) node[below] {$\scriptstyle{-r_-}$};
\draw[thick,color=red,fill=red] (-4.51413,0) circle (0.25ex) node[below] {$\scriptstyle{r_-}$};
\draw[thick,color=red,fill=red] (-5,2.158) circle (0.25ex) node[above] {$\scriptstyle{i r_i}$};
\draw[thick,color=red,fill=red] (-5,-2.158) circle (0.25ex) node[below] {$\scriptstyle{-i r_i}$};
\draw[thick,color=red,fill=red] (1,0) circle (0.25ex) node[below right] {$\scriptstyle{\infty}$};
\draw[thick,snake it, color=blue] (-3,0) -- (1,0) ;
\draw[thick,color=black, ->-] (-3,-0.3) -- (1,-0.3);
\draw[thick,color=black,->-] (1,0.3) -- (-3,0.3);
\draw[thick,color=black,-<-] (-3,-0.3) arc (315:45:0.424);
\end{tikzpicture}
\caption{ Singular points  of the designer scalar equation in \RNAdS{5} in the complex $r$ plane. The roots of $f$ are $\pm \{r_+, r_-, i\,r_i\}$, with
the inner horizon at $r_- = r_+ \sqrt{-\frac{1}{2} + \sqrt{\frac{1}{4} +Q^2}}$, and $r_i = \sqrt{\frac{1}{2} + \sqrt{\frac{1}{4} +Q^2}}$. These are depicted here for  $Q =\frac{1}{4}$, as is a circle of radius $r_-$ (faint dotted curve). The conventions are as in \cref{fig:designerSchw}, except that the origin which sometimes is an ASP is distinguished. }
\label{fig:designerRN}
\end{figure}

Since the \RNAdS{d+1} black hole has a different causal structure, in particular a timelike singularity, the behaviour at the curvature SP is different. In particular, the characteristic exponents are $0$ and $2d-3-\ann$, respectively. For a minimally coupled massive scalar, with $\ann=d-1$, not only are both exponents are integral, but the logarithmic branch can also be checked to be absent. Thus, in this case, we have a curvature ASP, with the fields not really being sensitive to the black hole singularity -- however, some components of the conserved currents have duals that are cognizant of the singularity, cf., \cref{sec:vRNads}. We believe this should hold for a range of $\ann$, but have not checked in detail whether it does. One can again check that the designer scalar wave equation in \RNAdS{d+1} background is a Fuchsian SOLDE, with $2d$ SPs (we include the curvature SP in this count even when it is an ASP). Essential features of this equation are summarized in \cref{fig:designerRN}.

\paragraph{Extremal confluence:} There are other additional features of the \RNAdS{d+1} solution that have to do with degenerate horizons, and mergers of zeros of $f(r)$. For an extremal black hole when the inner and outer horizon radii are equal\footnote{ 
    Note, however, that the proper distance between the inner and outer horizon remains finite in the limit.}
the horizon singular point becomes irregular due to a confluence of two regular singular points (say at $r=r_+$ and $r=r_-$).\footnote{
    If we play formal games of working with $Q\in \mathbb{C}$ then we can find scenarios where the complex roots of $f(r)$ coalesce. A simple example seems to be $d=4$ and $Q = \frac{i}{2}$, for which the roots are at $\pm r_+$, $\pm i \, \frac{r_+}{\sqrt{2}}$, each of the latter two being two-fold degenerate. }
We will not analyze this limit any further. 

We next turn to the vector and scalar perturbations of the metric and gauge field in Einstein-Maxwell theory about a \RNAdS{d+1} background. The equations have been compiled in \cref{sec:rnadseqs}.

%~~~~~~~~~~~~~~~~~~~~~~~~~~~~~~~~~~~~~~~~~~~~~~~
\subsubsection{Momentum diffusion ODE}
\label{sec:vRNads}
%~~~~~~~~~~~~~~~~~~~~~~~~~~~~~~~~~~~~~~~~~~~~~~ 

In the vector sector we have two equations, one for the momentum diffusion mode, $\MX$ and the other for the transverse photons $\MY$, cf.,~\eqref{eq:RNvector}. The former is of the form of a designer scalar with an exponent $\ann = 1-d$ but with a more involved dependence on the spatial momentum through $\BQTv$ defined in~\eqref{eq:BQTvdef}. We denote the coefficient functions appearing in the standard form of the SOLDE as $\{p_\MX, q_\MX\}$ and $\{p_\MY,q_\MY\}$, respectively.

The generic SPs of these equations have the following behaviour:
\begin{itemize}[wide,left=0pt]
\item  Asymptotic SP: $\MX$ behaves as a designer field with index $\ann= 1-d$, while $\MY$ behaves as one with index $d-3$. The potentials do not affect this analysis (but enter in the determination of counterterms).
\item Curvature SP: In the vicinity of $r=0$ we find
\begin{equation}\label{vRNcurvSP}
\begin{aligned}
p_\MX &= \frac{5-3\,d}{r} +\order{r^{d-3}}, &  
 q_\MX &= \order{r^{d-4}} ,\\
p_\MY &= \frac{5-3\,d}{r} + \order{r^{d-3}} , &
q_\MY
    &=  \order{r^{d-4}}.
\end{aligned}
\end{equation}
Since the coefficient functions approach the origin with power laws smaller than the difference in characteristic exponents (which are $0$ and $3d-4$, respectively) there is potentially logarithmic branch cut, whose presence can be confirmed by direct computation. Hence, for these fields the black hole singularity is a genuine SP, unlike the case for the other equations we analyze in the \RNAdS{d+1} background.
\item Horizon SP: At the horizon, the functions $p_\MX$ and $p_\MY$ have a simple pole with residue as in~\eqref{eq:DMhorSPRN}. The functions $q_\MX$ and $q_\MY$ also have simple poles, though the residue is a more complicated function of momentum 
\begin{equation}\label{eq:qvRNhorSP}
\begin{split}
q_\MX 
&=
    -\frac{\beta}{4\pi}\, \frac{1}{r-r_+} \left(\left( 1-\sdc^2 + \frac{1}{2}\, \BQTv^2\right) \bRQ^2\, \BQTv^2 - (d-1) \, i \bwt\right) + \cdots \,,\\
q_\MY 
&=
    - \frac{\beta}{4\pi}\, \frac{1}{r-r_+} \left(\left( 1+\sdc^2 + \frac{1}{2}\, \BQTv^2\right) \bRQ^2\, h(r_+)\, \BQTv^2 - \frac{d-3+(d-1)\,\sdc}{\sdc-1}\,  i \bwt\right) + \cdots \,.\\   
\end{split}
\end{equation}  
\item Ohmic SP: For the field $\MY$ owing to the dependence on the function $h$, we have additional singular points at 
\begin{equation}\label{eq:ohmSPloc}
r = \RQ\, \varpi_{d-2} \,.
\end{equation}  
Since the parameter $\RQ \in [0,r_+]$ with the lower bound attained on the neutral solution and the upper bound at extremality, these loci are always within the unit circle of radius $r_+$ where the primary horizon SP is located. The local behaviour in the vicinity of this SP (making the obvious phase choice $\varpi_{d-2}=1$ for simplicity) is 
\begin{equation}\label{eq:ohmSP}
p_\MY = \frac{2}{r-\RQ} +\order{(r-\RQ)^0}\,, \qquad q_\MY = -\frac{2\, r_+}{\RQ^2\, f(\RQ)}\, \frac{i\bwt}{r-\RQ} +\order{(r-\RQ)^0}\,.
\end{equation}  
\end{itemize}

\begin{figure}[ht!]
\centering
\begin{tikzpicture}[scale=1]
\draw[very thin,color=black,dashed] (-5,0) circle (2cm);
\draw[very thin,color=black,dotted] (-5,0) circle (0.54858cm);
\draw[very thin,color=black,dotted] (-5,0) circle (1cm);
\draw[thick,color=red,fill=red] (-5,0) circle (0.25ex) node[below] {$\scriptstyle{0}$};
\draw[thick,color=red,fill=red] (-7,0) circle (0.25ex) node[left] {$\scriptstyle{-r_+}$};
\draw[thick,color=red,fill=red] (-3,0) circle (0.25ex) node[left] {$\scriptstyle{r_+}$};
\draw[thick,color=red,fill=red] (-5.54858,0) circle (0.25ex) node[above] {$\scriptstyle{-r_-}$};
\draw[thick,color=red,fill=red] (-4.51413,0) circle (0.25ex) node[above] {$\scriptstyle{r_-}$};
\draw[thick,color=red,fill=red] (-5,2.158) circle (0.25ex) node[above] {$\scriptstyle{i r_i}$};
\draw[thick,color=red,fill=red] (-5,-2.158) circle (0.25ex) node[below] {$\scriptstyle{-i r_i}$};
\draw[thick,color=red,fill=red] (-4,0) circle (0.25ex) node[below] {$\scriptstyle{\RQ}$};
\draw[thick,color=red,fill=red] (-6,0) circle (0.25ex) node[below] {$\scriptstyle{-\RQ}$};
\draw[thick,color=red,fill=red] (1,0) circle (0.25ex) node[below right] {$\scriptstyle{\infty}$};
\draw[thick,snake it, color=blue] (-3,0) -- (1,0) ;
\draw[thick,color=black, ->-] (-3,-0.3) -- (1,-0.3);
\draw[thick,color=black,->-] (1,0.3) -- (-3,0.3);
\draw[thick,color=black,-<-] (-3,-0.3) arc (315:45:0.424);
\end{tikzpicture}
\caption{ Singular points  of the transverse photon equation in \RNAdS{5} in the complex $r$ plane. The conventions are as in \cref{fig:designerRN}; we have now made clear that the origin is a  regular SP, and indicated the Ohmic SPs for this equation as well. Note that these lie between the inner and outer horizons~\cite{He:2021jna}.  }
\label{fig:YRN}
\end{figure}
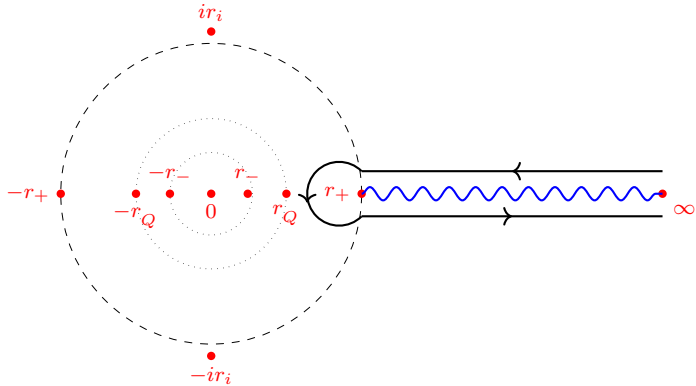

The black hole singularity is a genuine regular SP for both $\MX$ and $\MY$ unlike the designer scalars examined above. This is despite the characteristic exponents being integral, owing to the presence of a logarithmic branch. This is, per se, curious, since the behaviour of these fields dual to linear combination of conserved current operators appears to subvert the expectations one might have had for timelike singularities. Note that the equations are indeed Fuchsian since the sum of the characteristic exponents does satisfy the Fuchsian condition. $\MX$ has $2\,d$ SPs, while $\MY$ as $3\,d-2$ SPs. Thus, the equation for $\MX$ behaves like the designer scalar system depicted in \cref{fig:designerRN} (with the origin being a regular SP), while $\MY$ had additional SPs which are now depicted in \cref{fig:YRN}.

Apart from the generic behaviour described above, we have also the possibility of extremal confluence and Matsubara ASPs. For $\MX$, the extremal confluence is similar to that encountered for generic designer scalars. However, for $\MY$ there is the additional complication since $\RQ \to r_+$ in this limit. It turns out that the difference is that while $q_\MX$ has a second order pole in the extremal limit, $q_\MY$ has a third order pole (both $p_\MX$ and $p_\MY$ have a double pole). The behaviour at the Matsubara ASP is no different from the earlier discussion -- the only change is that the values of momenta where both modes are analytic is determined by a polynomial equation for 
$\BQTv^2$.  

%~~~~~~~~~~~~~~~~~~~~~~~~~~~~~~~~~~~~~~~~~~~~~~~
\subsubsection{Energy density and charge diffusion ODEs}
\label{sec:sRNads}
%~~~~~~~~~~~~~~~~~~~~~~~~~~~~~~~~~~~~~~~~~~~~~~ 

Let us turn to the scalar sector where there are some novel features. The equations of motion are given in~\eqref{eq:RNscalar} and have quite complicated expressions for the potentials. Furthermore, the dependence on the spatial momentum again appears in a surdic form through a parameter $\BQTs$ defined in~\eqref{eq:BQTsdef}, though in a manner different from that in the case of the  vector perturbations. It turns out to be useful work with $\BQTs$ directly, eliminating $\bqt$ (or $k$). To do so, we recall a useful identity from~\cite{He:2022deg}
\begin{equation}\label{eq:BQTsrel}
\BQTs^2 \, (\BQTs^2+2) = \frac{2\, d\, \nu_s}{\bRQ^2} \, \bqt^2 \,,
\end{equation}  
which follows from the definition. This can be used to simplify the function $\Lk$ (whose functional form is identical in terms of the metric function as for the \SAdS{d+1} case~\eqref{eq:Lkdef}) as follows:
\begin{equation}
\Lk = \frac{(d-1)^2\, Q^2}{4\, \sdc^2} \left(2+\BQTs^2 - 2\,\frac{\RQ^{d-2}}{r^{d-2}}\right)\left(\BQTs^2 +2\, \frac{\RQ^{d-2}}{r^{d-2}}\right) . 
\end{equation}  
We are now ready to describe the singular points of the SOLDEs for $\Zd$ and $\Vd$. 

\begin{itemize}[wide,left=0pt]
\item  Asymptotic SP: Both $\Vd$ and $\Zd$ behave as designer fields with index $3-d$. The potentials again are irrelevant for this analysis, and these fields behave as non-Markovian modes with characteristic exponents of $0$ and $d-4$.
\item Curvature ASP: The black hole singularity continues to be a singular point of the $\Zd$ and $\Vd$ SOLDE. The coefficient functions at most have simple poles. Specifically, 
\begin{equation}\label{eq:RNZsing}
\begin{split}
p_\Zd(r) &= 
    - \frac{d -3}{r} + \order{r^{d-3}} \,, \qquad 
    q_\Zd(r) = \order{r^{2\,(d-3)}} , \\ 
p_\Vd(r) 
&= 
    - \frac{d -3}{r} + \order{r^{d-3}} \,, \qquad 
    q_\Vd(r) = \order{r^{2\,(d-3)}}  . 
\end{split}
\end{equation}  
The characteristic exponents are integral and there is no logarithmic branch cut in the solutions. Hence, black hole singularity is an ASP of the equations for both $\Zd$ and $\Vd$.
\item Horizon SP: The behaviour of $p_\Vd$ and $p_\Zd$ is as for any other field in a non-extremal background; a simple pole with residue $1-i\mathbf{w}$. The functions $q_\Vd$ and $q_\Zd$ also only have simple poles, but with a residue that is depends non-trivially on $\BQTs$ and $\sdc$ (or $Q$). 
\item Ohmic SP: The fields $\Vd$ and $\Zd$ have Ohmic SP, since their kinetic terms are modulated by $h$. The functions $p_\Vd$ and $p_\Zd$ have a simple poles with residue $\mp2$, respectively, while $q_\Vd$ and $q_\Zd$ start off with a double pole, as can be read-off directly from~\eqref{eq:pqVd} and~\eqref{eq:pqZd}. 
\end{itemize}

\paragraph{Energy density (A)SP:} In the \SAdS{d+1} analysis, we saw that the roots of $\Lk$, which naively appear to be SPs of the SOLDEs, are actually not. We will confirm this to be the case for the \RNAdS{d+1} equations as well, albeit in a slightly complicated manner. Notice that the vanishing loci of $\Lk$ are located at the following $2(d-2)$ points of the complex radial plane. 
\begin{equation}\label{eq:Lkzeros}
r_{k1}= \RQ \left(-\frac{\BQTs^2}{2} \right)^{-\frac{1}{d-2}} \, \varpi_{d-2}
\,,
\qquad 
r_{k2}= \RQ \left(\frac{2 +\BQTs^2}{2} \right)^{-\frac{1}{d-2}}\, \varpi_{d-2}\,. 
\end{equation}  
The first set here is continuously connected to the locations where the corresponding function for the neutral black hole has roots, while the second set is unique to the charged black hole. 

To deduce this, we ask which of the two roots can be made to lie on the ray $[r_+, \infty)$ along the real axis by choosing an appropriate phase for $\BQTs$. Since $\BQTs^2 \sim \bqt^2$ for small momentum it follows that the first set of roots connect to the neutral case. Therefore, for small $\bqt$ the first set of roots lies outside the circle $\abs{r} =r_+$, while the second set lies within. At large charge however, the situation is reversed, with the exchange between them occurring at a critical charge
\begin{equation}\label{eq:sdccritical}
\sdc^* = \frac{1}{2} \,, \qquad Q^2_* = \frac{d}{3\,d-4}\,.
\end{equation}  
When $\sdc <\frac{1}{2}$, we need to ascertain whether the first set of  zeros of $\Lk$ lies between the horizon and the boundary, but the question switches over to the second branch for larger charges, with $\sdc=\frac{1}{2}$ being a critical point.  For the present however, we can simply analyze the two sets of roots in turn, not worrying about whether they lie within or outside the circle of radius $r_+$.  

Let us start with the particular root $r_{k1}= \RQ \left(-\frac{\BQTs^2}{2} \right)^{-\frac{1}{d-2}}$ (fixing the phase). For the field $\Zd$ we find that the function $p_\Zd$ has a simple pole with residue $-2$. The function $q_\Zd$ also has only a simple pole, leading to characteristic exponents $0$ and $3$, respectively. We then expand the functions to linear order around this putative SP to check if there is a potential logarithmic term, and again find that it is absent, rendering the singular point apparent. The analysis for $\Vd$ is a bit simpler. At this zero of $\Lk$ one can check that the functions $\{p_\Vd,q_\Vd\}$ are, in fact, manifestly regular, thus rendering the zero set to be ordinary points of the SOLDE.

When we pick the root $r_{k2}= \RQ \left(1+\frac{\BQTs^2}{2} \right)^{-\frac{1}{d-2}}$ (with phase fixed to unity), we find that  $p_\Zd$ has a simple pole with residue $-2$, while $q_\Zd$ has a double pole with residue $2$. The characteristic exponents are $1$ and $2$, respectively. Examining the constant term in $p_\Zd$ we find that it is negative of the residue of $q_\Zd$ at its simple pole term. This guarantees the absence of any monodromy around this SP, rendering it also to be an apparent SP. For the field $\Vd$ however, we find a double pole in $q_\Vd$ at this root, with residue $-2$. This implies that the characteristic exponents are $-1$ and $2$. We have checked that there is no monodromy around this point as well. However, the solution for the field $\Vd$ does have a simple pole at these roots of $\Lk$.  Ergo, the set of roots $r_{k2}$ is a genuine SP of the SOLDE.

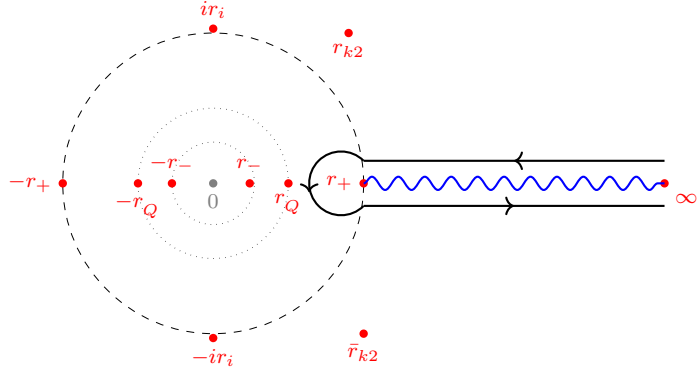
\begin{figure}[ht!]
\centering
\begin{tikzpicture}[scale=1]
\draw[very thin,color=black,dashed] (-5,0) circle (2cm);
\draw[very thin,color=black,dotted] (-5,0) circle (0.54858cm);
\draw[very thin,color=black,dotted] (-5,0) circle (1cm);
\draw[thick,color=gray,fill=gray] (-5,0) circle (0.25ex) node[below] {$\scriptstyle{0}$};
\draw[thick,color=red,fill=red] (-7,0) circle (0.25ex) node[left] {$\scriptstyle{-r_+}$};
\draw[thick,color=red,fill=red] (-3,0) circle (0.25ex) node[left] {$\scriptstyle{r_+}$};
\draw[thick,color=red,fill=red] (-5.54858,0) circle (0.25ex) node[above] {$\scriptstyle{-r_-}$};
\draw[thick,color=red,fill=red] (-4.51413,0) circle (0.25ex) node[above] {$\scriptstyle{r_-}$};
\draw[thick,color=red,fill=red] (-5,2.058) circle (0.25ex) node[above] {$\scriptstyle{i r_i}$};
\draw[thick,color=red,fill=red] (-5,-2.058) circle (0.25ex) node[below] {$\scriptstyle{-i r_i}$};
\draw[thick,color=red,fill=red] (-4,0) circle (0.25ex) node[below] {$\scriptstyle{\RQ}$};
\draw[thick,color=red,fill=red] (-6,0) circle (0.25ex) node[below] {$\scriptstyle{-\RQ}$};
\draw[thick,color=red,fill=red] (-3.2,2) circle (0.25ex) node[below] {$\scriptstyle{r_{k2}}$};
\draw[thick,color=red,fill=red] (-3,-2) circle (0.25ex) node[below] {$\scriptstyle{\bar{r}_{k2}}$};
\draw[thick,color=red,fill=red] (1,0) circle (0.25ex) node[below right] {$\scriptstyle{\infty}$};
\draw[thick,snake it, color=blue] (-3,0) -- (1,0) ;
\draw[thick,color=black, ->-] (-3,-0.3) -- (1,-0.3);
\draw[thick,color=black,->-] (1,0.3) -- (-3,0.3);
\draw[thick,color=black,-<-] (-3,-0.3) arc (315:45:0.424);
\end{tikzpicture}
\caption{ Singular points  of the longitudinal scalar mode $\Vd$, which is dual to the charge diffusion mode photon in \RNAdS{5} in the complex $r$ plane. The conventions are as in \cref{fig:YRN}. The new elements here are the energy density SPs from the $d-2$ roots of $\Lk$ in the set $r_{k2}$, which we have labeled as $\{r_{k2}, \bar{r}_{k2}\}$. For the latter, we have chosen non-zero value for  $\arg(k^2)$ to aid with the depiction, but note that as we rotate $k$ we can make the SP pinch the grSK contour.  }
\label{fig:VRN}
\end{figure}

The above completes the description of the singularity structure for generic values of parameters for non-extremal \RNAdS{d+1} black holes. There are again features specific to fine-tune loci in parameter space. Some of these are as before, viz., 
\begin{itemize}[wide,left=0pt]
\item Extremal confluence: When we take the extremal limit, demanding that $f$ has a double zero at $r=r_+$, we convert $r=r_+$ to an irregular SP of the equations. 
\item Energy asymptotic coalescence:   As in the \SAdS{d+1} case we have a special situation when the spatial momentum vanishes for the energy density mode $\Zd$. Owing to the presence of $\Lk$ in the kinetic term, at $k^2=0$, we find coalescence of the roots of $\Lk$ which are now at the asymptotic boundary. This leaves the asymptotic singular point regular, but changes the characteristic exponents to $1$ and $-d$, respectively. Note that this behaviour is not present for the charge diffusion field $\Vd$, since its kinetic term does not have a corresponding factor of $\Lk$. 
\item Matsubara ASP: As in previous equations we can tune to specific frequencies $i\mathbf{w}= n$ with $n \in \mathbb{Z}_{\geq 0}$. At these values the outgoing mode can be made regular, provided we remove the monodromy due to the logarithmic branch of the solution in the outgoing mode.
To achieve this we have to tune $k^2$, or equivalently $\BQTs$, which we have used to parameterize both the $\Zd$ and $\Vd$ equations.  
\end{itemize}

\paragraph{Energy horizon coalescence:} Now that we have understood the behaviour at the horizon, and at the roots of $\Lk$, let us turn to the case where we tune the momenta for the singular points arising from these two to merge. This will happen whenever we have tuned the momenta to be such that the root of $\Lk$ lies at $r=r_+$. It is easy enough to determine when this happens, we need either 
\begin{equation}\label{eq:LkhormergerBQT}
\BQTs^2 =
\begin{cases}
-2\,\sdc \,, \qquad 0 < \sdc < \frac{1}{2} \,, \\
2\, (\sdc-1) \,, \qquad  \frac{1}{2} < \sdc < 1\,,
\end{cases}
\end{equation}  
based on the two branches of zeros. One can check that this is consistent with fixing the physical momentum using 
\begin{equation}\label{eq:Lkhormergerk}
k^2 = -\frac{d-1}{2}\, r_+^3\, f'(r_+)\,.
\end{equation}  
At this point it becomes clear that the jump is due to the fact that there is branch structure in the momentum dependence, owing to the presence of $\BQTs$. 

Let us start with the energy density equation for $\Zd$ and examine the two roots in~\eqref{eq:LkhormergerBQT} in turn. When $\BQTs^2 = -2\,\sdc $ we find the local behaviour of the coefficient functions to be
\begin{equation}\label{eq:pqZRNmergerRoot1}
\begin{split}
p_\Zd
&= 
    -\frac{1+i \mathbf{w}}{r-r_+} + \order{(r-r_+)^0} \,,\\
q_\Zd
&= 
    \frac{1+i \mathbf{w}}{(r-r_+)^2} + \frac{d-5 + (17-4\,d)\,\sdc+6\,(d-3)\,\sdc^2}{2\,(2\,\sdc-1)\,(\sdc-1)}\, \frac{1+i \mathbf{w}}{r-r_+}+ \order{(r-r_+)^0}  .
\end{split}
\end{equation}  
On the other hand for the root $\BQTs^2 = 2\,(\sdc-1) $ we find
\begin{equation}\label{eq:pqZRNmergerRoot2}
\begin{split}
p_\Zd
&= 
    -\frac{1+i \mathbf{w}}{r-r_+} + \order{(r-r_+)^0} \,,\\
q_\Zd
&= 
    \frac{1+i \mathbf{w}}{(r-r_+)^2} \\
&\quad   
     + 
        \frac{3 \left(d-3 + (7-2\,d)\,\sdc+2\,(d-3)\,\sdc^2\right) + 
     \left(d-5 + (17-4\,d)\,\sdc+6\,(d-3)\,\sdc^2\right) \,i \mathbf{w}}{2\,(2\,\sdc-1)\,(\sdc-1)\, (r-r_+)} \\
&\qquad
    + \order{(r-r_+)^0} .
\end{split}
\end{equation}  

From both roots we see that the coalescence leaves the horizon locus as a regular SP with characteristic exponents, $1$ and $1+i \mathbf{w}$, exactly as in the \SAdS{d+1} case analyzed earlier in \cref{sec:energyLk}. However, now we find that only for the root $\BQTs^2 = -2\,\sdc$ does one encounter a value of the frequency, $\omega = 2\pi i\,T$, for which the horizon is an apparent singularity. For the second root this fails to be the case, since the non-vanishing of the residue of the simple pole at this frequency. In fact, at this frequency we find 
\begin{equation}\label{eq:ResZqhorR2}
\Res[q_\Zd]_{r=r_+,  \mathbf{w}=i} = \frac{d-2}{2\,\sdc-1} \,,
\end{equation}  
making it clear that there is a potential logarithmic piece in the solution, leading to a non-trivial monodromy.

Let us translate this observation into physical parameters. The question boils down to the constraint on the charge of the \RNAdS{d+1} black hole which allows for the merger of the root of $\Lk$ with the horizon locus. This occurs for the root $\BQTs^2 = -2\,\sdc$ for small charges, but for the other root, $\BQTs^2  = 2\,(\sdc-1)$ for larger values of charges. The separatrix is characterized by the critical charge~\eqref{eq:sdccritical}. 

Thus, for $\sdc \in[0,\frac{1}{2})$, we learn that the horizon coalescence leads to an ASP when $k^2$ is tuned, for a particular frequency, $\omega = 2\pi i\, T$. However, for $\sdc \in (\frac{1}{2},1)$ the horizon remains a regular SP, and there is no interesting feature in the equation at this special value of the frequency. There is something special at $\sdc = \frac{1}{2}$, $\Lk(r)$ has a double-zero at the horizon locus,
\begin{equation}\label{eq:Lkstar}
\Lk(r)= \frac{d\,(d-1)^2}{3d-4} \left(1-\frac{r_+^{d-2}}{r^{d-2}}\right)^2 \,, \qquad k^2 = - \frac{d-1}{2}\, r_+^3\, f'(r_+)\,.
\end{equation}  
While this could have led to the horizon becoming an irregular SP, strangely enough, it doesn't  and remains a regular singular point. However, at this special point in parameter space, the residues are again modified,
\begin{equation}\label{eq:pqZRNmergerstar}
\begin{split}
p_\Zd
&= 
    -\frac{3+i \mathbf{w}}{r-r_+} + \order{(r-r_+)^0} \,,\\
q_\Zd
&= 
    2\,\frac{2+i \mathbf{w}}{(r-r_+)^2} + \frac{7-9\,d - (3\,d-1)\,i \mathbf{w}}{r-r_+}+ \order{(r-r_+)^0}  . 
\end{split}
\end{equation}  
The characteristic exponents are $\alpha_1=2$ and $\alpha_2 = 2 + i \mathbf{w}$. It is easy to check that the  singular point always remains regular, and the outgoing solution, non-analytic.\footnote{
    The order of limits is important here -- we first fix $\sdc=\frac{1}{2}$ and then tune the zero of $\Lk$ by fixing the momentum.  
}

One can understand this behaviour physically along the following: we know that for the \SAdS{d+1} black hole the horizon coalescence is an ASP for the smallest positive Matsubara frequency $\omega = 2\pi i\,T$. By continuity, we expect this to be the case for the \RNAdS{d+1} perturbations for small values of the charge. This can be easily verified by checking that the $\Zd$ equation in~\eqref{eq:RNscalar} reduces nicely to the $\MZ$ equation as $Q \to 0$. Thus, on physical grounds there is an open neighbourhood of $Q=0$ where the \RNAdS{d+1} energy density mode ought to exhibit similar behaviour. The question then is whether this behaviour ceases at some finite value of the charge, or extend all the way to the extremal limit $Q^2 = \frac{d}{d-2} $ or $\sdc =1$. We just learned that the physical boundary is democratic in the parameter $\sdc$, which measures the Ohmic radius $\RQ$ in units of the horizon size,~\eqref{eq:sdcdef}. 

So what happens above this critical value of the charge? It can't be that the equations completely forget about the coalescence, which would only make sense if the roots of the $\Lk$ stopped merging, but that as we see is not the case. In fact, what happens is a transference of behaviour to the charge density mode! 

Let's therefore examine the equation for $\Vd$ at these roots of $\Lk$ when the merge with the horizon. For the first root set $r_{k1}$ obtained  by setting $\BQTs^2 = -2\,\sdc$ which is valid for $\sdc \leq \frac{1}{2}$ we find:
\begin{equation}\label{eq:pqVRNmergerRoot1}
\begin{split}
p_\Vd
&= 
    \frac{1-i \mathbf{w}}{r-r_+} + \order{(r-r_+)^0} \,,\\
q_\Vd
&= 
    \left[ \frac{(d-1)\,(1+2\,\sdc^2) -(d+1)\,\sdc }{2\, (\sdc-1)\, (2\,\sdc-1)} - \frac{d-3 +(d-1)\,\sdc}{2\, (\sdc-1)}\, i\mathbf{w}  \right]\frac{1}{r-r_+} \\
&\qquad 
    + \order{(r-r_+)^0} \,.
\end{split}
\end{equation}  
Now the indicial equation leads to characteristic exponents $0$ and $i \mathbf{w}$, which may lead to analytic solutions only for negative Matsubara frequencies. Furthermore, for any choice of $\mathbf{w}= -i\,n$ with $n\in \mathbb{N}$, we see that the residue of the simple pole of $q_\Vd$ is non-vanishing. For example, 
\begin{equation}\label{eq:ResVqhorR1}
\Res[q_\Vd]_{r=r_+,  \mathbf{w}=-i} = -\frac{d-2}{2\,\sdc-1} \,.
\end{equation}  
Thus, the horizon remains a regular SP of the $\Vd$ equation when the coalescence happens with the roots in the $r_{k1}$ set. 

However, at the second set of roots $r_{k2}$ of $\Lk$, viz., $\BQTs^2 = 2\,(\sdc-1)$ which merges with the horizon for $\sdc > \frac{1}{2}$ we find upon coalescence
\begin{equation}\label{eq:pqVRNmergerRoot2}
\begin{split}
p_\Vd
&= 
    \frac{1-i \mathbf{w}}{r-r_+} + \order{(r-r_+)^0} \,,\\
q_\Vd
&= 
     \frac{d-3 +(d-1)\,\sdc}{2\, (\sdc-1)}\,\frac{1- i \mathbf{w}}{r-r_+} 
    + \order{(r-r_+)^0} \,.
\end{split}
\end{equation}  
Now, for $\mathbf{w}= -i$ we see that the residue of the simple 
pole in $q_\Vd$ vanishes, allowing for this coalescence to turn itself into an apparent singularity.

Finally, at $\sdc= \frac{1}{2}$, we simply encounter a regular SP, despite the double-root of $\Lk(r)$, as noted in~\eqref{eq:Lkstar}. Now we find
\begin{equation}\label{eq:pqVRNmergerstar}
\begin{split}
p_\Vd
&= 
    \frac{1-i \mathbf{w}}{r-r_+} + \order{(r-r_+)^0} \,,\\
q_\Vd
&= 
  -\frac{(3\,d - 7)\,(1-i \mathbf{w})}{r-r_+}+ \order{(r-r_+)^0}  . 
\end{split}
\end{equation}  
For $\mathbf{w}=-i$ we again find an apparent singularity --  in fact, this behaviour seems to smoothly connect to the second branch of $\Lk$ roots unlike what happened for the $\Zd$ equation above.

%~~~~~~~~~~~~~~~~~~~~~~~~~~~~~~~~~~~~~~~~~~~~~~~
\section{Form factors in the FPF correlator}\label{sec:Jfns}
%~~~~~~~~~~~~~~~~~~~~~~~~~~~~~~~~~~~~~~~~~~~~~~

We collect in this appendix the formulae for the functions $ \mathfrak{J}_{\text{reg}}$ and $ \mathfrak{J}_{\text{loc}}$, which enter into our analysis of the three-point function evaluated in~\cref{sec:example}.

Let us start with $ \mathfrak{J}_{\text{reg}}$ entered the regular piece of the three-point function. This function involves completing the Mellin-Barnes contour integrals, which, we recall, were introduced to allow one to complete the radial integral over a product of hypergeometric functions. There are three propagators, and hence, three contour integrals. These are evaluated by standard residue calculus, and lead to the following infinite sum representation
\begin{equation}
\mathfrak{J}_{\text{reg}} =  \frac{1}{8\,(\Delta_\phi-1)^2\,(\Delta_\chi-1)}
 \sum_{\delta_i=\{\Delta_i,2-\Delta_i\}} \,
    \sum_{\vec{n}=1}^{\infty}
    \, \mathcal{J}_{\text{reg}}^{\vec{\delta}}(\vec{n})
\end{equation}  
The summations over $\vec{n} =\{n_1,n_2,n_3\}$ runs over the natural numbers, and arises from closing the Mellin-Barnes contours. In the process, the residues get contributions from both the primaries of dimension $\Delta_i$, and the shadow dimension $2-\Delta_i$. We recall that $\Delta_1=\Delta_2 = \Delta_\phi$, and $\Delta_3= \Delta_\chi$, for brevity. The expression for the summand itself is 
\begin{equation}\label{eq:Jformreg}
\begin{split}
\mathcal{J}_{_\text{reg}}^{\vec{\delta}}(\vec{n}) 
&=
    \left(\prod_{i=1}^{3}
        \frac{(-1)^{n_i}}{\Gamma(1+n_i)}
        \frac{\Gamma(1-\delta_i-n_i)}{\Gamma(1-\delta_i-2n_i)}\right)\, 
    \frac{\gfn{}(k_1,\delta_1+2n_1)}{\gfn{}(k_1,\Delta_{\phi})}
    \frac{\gfn{}(\bar{k}_2,\delta_2+2n_2)}{\gfn{}(\bar{k}_2,\Delta_{\phi})}
    \frac{\gfn{}(k_3,\delta_3+2n_3)}{\gfn{}(k_3,\Delta_{\chi})}
    \mathfrak{h}_{\vec{n}}\,, \\ 
\mathfrak{h}_{\vec{n}} 
&=
    \bqt_1\cdot\bqt_2 \,\mathfrak{H}_{1,0} -\left(3\,\bwt_1\,\bwt_2+(\delta_1+2n_1)\,(\delta_2+2n_2)\right)\mathfrak{H}_{1,1} \\ 
&\quad 
    -\left(\bwt_1\, \bwt_2+i\bwt_2\,(\delta_1+2n_1)-i\bwt_1\,(\delta_2+2n_2)+(\delta_1+2n_1)\,(\delta_2+2n_2)\right) \mathfrak{H}_{0,1}\\ 
&\quad 
    +\left(i\bwt_2\,(\delta_1+2n_1)-i\bwt_1\,(\delta_2+2n_2)+(\delta_1+2n_1)\,(\delta_2+2n_2)\right) \mathfrak{H}_{2,1} \\ 
&\quad 
    +(\delta_1+2n_1)\, (\delta_2+2n_2)\, \mathfrak{H}_{3,1} 
    + 2i\left(\bwt_1\, (\delta_2+2n_2)-\bwt_2\, (\delta_1+2n_1)\right)
    \mathfrak{H}_{2,0}\,,\\
\mathfrak{H}_{a,b} 
&=
    H_{\bwt_2}\left(\sum_{i}(\delta_i+2n_i)-a,-b\right)\,.
\end{split}
\end{equation}
We have introduced here the function $H_{\bwt}(a,b)$, which arises from performing the radial integrals. It is defined to be 
\begin{equation}
H_{\bwt}(a,b) \equiv \frac{2}{\Gamma\left(1+i\bwt\right)}\int_{0}^{1} dz \left(1-z^2\right)^{i\bwt}z^a(1+z)^b\,,
\end{equation}
with the prefactor chosen for convenience. The function is well-defined provided that $\text{Re}(a)>-1$, and we can evaluate it in closed form in terms of regularized hypergeometric functions. In fact, we only need the cases $b=0$ and $b=-1$, which turn out to have particularly simple expressions:
\begin{equation}
\begin{split}
H_{\bwt}(a,0) 
&= 
    \frac{\Gamma\left(\frac{1+a}{2}\right)}{\Gamma\left(i\bwt+\frac{a+3}{2}\right)}\,,\\ 
H_{\bwt}(a,-1) 
&= 
    \Gamma\left(\frac{1+a}{2}\right)
        \pFqReg{2}{1}{1,\frac{a+1}{2}}{i\bwt+\frac{a+3}{2}}{1}
    -\Gamma\left(\frac{2+a}{2}\right)
        \pFqReg{2}{1}{1,\frac{a+2}{2}}{i\bwt+\frac{a+4}{2}}{1}
\end{split}
\end{equation}

The evaluation of the localized contribution follows along similar lines, once we note the absence of IR divergences from the contour integral. At the end of the day, we find that 
\begin{equation}\label{eq:Jformlocal}
\begin{split}
\mathcal{J}_{_\text{loc}}^{\vec{\delta}}(\vec{n})
&=
     \frac{1}{8\,(\Delta_\phi-1)^2\,(\Delta_\chi-1)}\left(\prod_{i=1}^{3}\frac{(-1)^{n_i}}{\Gamma(1+n_i)}\frac{\Gamma(1-\delta_i-n_i)}{\Gamma\left(1-\delta_i-2n_i\right)}\right)\\ 
&\quad 
    \times \frac{\gfn{}(k_1,\delta_{1}+2n_1)}{\gfn{}(k_1,\Delta_{\phi})}\frac{\gfn{}(\bar{k}_2,\delta_{2}+2n_2)}{\gfn{}(\bar{k}_2,\Delta_{\phi})}
    \frac{\gfn{}(k_3,\delta_{3}+2n_3)}{\gfn{}(k_3,\Delta_{\chi})}
    \frac{\Gamma\left(\frac{1}{2}\sum_i(\delta_i+2n_i)\right)}{\Gamma\left(i\bwt_2+\frac{1}{2}\sum_i(\delta_i+2n_i)\right)}
\,.
\end{split}
\end{equation}  

Adding up the contributions we find
\begin{equation}\label{eq:Jformfinal}
\begin{split}
\mathfrak{J} 
&=\mathfrak{J}_{_\text{loc}} -4i\,\bwt_1\, \mathfrak{J}_{_\text{reg}}
=
    \frac{1}{8\,(\Delta_\phi-1)^2\,(\Delta_\chi-1)}\,  \sum_{\delta_i=\{\Delta_i,2-\Delta_i\}} \,
    \sum_{\vec{n}=1}^{\infty} \, \mathcal{J}^{\vec{\delta}}(\vec{n}) \\
\mathcal{J}^{\vec{\delta}}(\vec{n})
&=
    \left(\prod_{i=1}^{3}\, \frac{(-1)^{n_i}}{\Gamma(1+n_i)}\frac{\Gamma(1-\delta_i-n_i)}{\Gamma\left(1-\delta_i-2n_i\right)}\right)
    \frac{\gfn{}(k_1,\delta_{1}+2n_1)}{\gfn{}(k_1,\Delta_{\phi})}\frac{\gfn{}(\bar{k}_2,\delta_{2}+2n_2)}{\gfn{}(\bar{k}_2,\Delta_{\phi})}\frac{\gfn{}(k_3,\delta_{3}+2n_3)}{\gfn{}(k_3,\Delta_{\chi})}\\ 
&\quad 
\times 
    \left[\mathfrak{h}_{\vec{n}}-4i\bwt_1\frac{\Gamma\left(\frac{1}{2}\sum_i(\delta_i+2n_i)\right)}{\Gamma\left(i\bwt_2+\frac{1}{2}\sum_i(\delta_i+2n_i)\right)}\right]\,.
\end{split}
\end{equation}
The reader can check that this expression does not lead to any additional singularities of the FPF correlation function of the operators dual to $\phi$ and $\chi$, respectively.

%~~~~~~~~~~~~~~~~~~~~~~~~~~~~~~~~~~~~~~~~~~~~~~~
\section{Local analysis for apparent singularities}
\label{sec:localapparent}
%~~~~~~~~~~~~~~~~~~~~~~~~~~~~~~~~~~~~~~~~~~~~~~

Consider a regular singular point, whose characteristic exponents are integer separated. It is helpful to recast the equation, scaling out the smaller exponent, and write a local form of the equation. We will focus on a particular form inspired by the equations we want to study, and w.l.o.g.\ place the singular point the origin. The equation we study is therefore
\begin{equation}\label{eq:localform}
\dv[2]{y}{z} +\left[ \frac{1-n}{z} + \sum_{m=0}^\infty\, p_m\, z^m\right] \, \dv{y}{z} + \left[ \frac{q_{-1}}{z} + \sum_{m=0}^\infty\, q_m\, z^m\right]y(z) =0
\end{equation}  
We have expanded the coefficient functions in a Laurent series, fixing the polar terms based on the structure we encounter for Fuchsian equations. The characteristic solutions are $z^0$ and $z^n$, and so the general solution is 
\begin{equation}\label{eq:genlog}
y(z) = A y_n(z) + B (y_0(z) + \gamma\, y_n(z) \, \log z ) \,. 
\end{equation}  
Here $y_n(z)$ and $y_0(z)$ are the naive Frobenius solutions $y_n(z)  = z^n \sum_{j=0}^\infty a_j\, z^j$ and $y_0(z)  =  \sum_{j=0}^\infty b_j\, z^j$, respectively.  We wish to understand the conditions under which the log term can be made absent, or equivalently how the monodromy matrix set to be the identity matrix. An early analysis can be found in~\cite{Heun:1888log}.

One strategy is to truncate the Laurent expansions for the coefficient functions and attempt to solve the SOLDE; this leads to closed form solutions for $n=1$ and $n=2$. The former is reliable, but the latter misses out some admixture from higher order terms in the coefficient functions. A more straightforward approach is the following: setting $a_0=1$, fix $a_1$ by solving the SOLDE. Then use the data to solve for the second independent function, and extract the condition for $\gamma =0$, as a constraint on the series coefficients. Carrying out the exercise we find the following condition for the monodromy $\gamma$ to vanish: 
\begin{equation}\label{eq:nologcond}
\begin{split}
n=1: & \quad q_{-1} =0 \,,\\
n=2: & \quad q_{-1}^2 + p_0\, q_{-1} + q_0  =0 \,, \\
n=3: & \quad q_{-1}^3 +2\, p_0\, q_{-1}^2 + 2\,q_{-1}\,(p_0^2 + p_1+ 2\,q_0) + 4\, (p_0 \, q_0 + q_1)   =0
\end{split}
\end{equation}  
The larger the difference in the characteristic exponents, the higher one needs to go in the series solution. Correspondingly, one is sensitive to more terms in the Laurent expansion of the coefficient functions about the singular point. This point has been explained in~\cite{Blake:2019otz} for some specific  cases; they express the result as a vanishing condition for a determinant of Frobenius series coefficients. The analysis presented here can be adapted to any regular singular point.

%%%%%%%%%%%%%%%%%%%%%%%%%%%%%%%%%%%%%%%%%%%%%%%

\providecommand{\href}[2]{#2}\begingroup\raggedright\endgroup

% \bibliographystyle{jhep}
% \bibliography{odecurrents}

\end{document}